\documentclass[aps,
		pra,
		reprint,
		twocolumn,
		superscriptaddress,
		shortbibliography,
		nofootinbib,
		floatfix,
		notitlepage
		]{revtex4-2}
\pdfoutput=1

\usepackage{graphicx}
\usepackage{mathrsfs}
\usepackage{bbm}
\usepackage{bbold}
\usepackage{amsfonts,amssymb,amsmath}
\usepackage{slashed}
\usepackage[]{units}
\usepackage[dvipsnames]{xcolor}
\usepackage{feynmp-auto}
\usepackage[export]{adjustbox}
\usepackage{hyperref}

\newcommand{\eps}{\varepsilon}

\newcommand{\vphi}{\varphi}
\newcommand{\vkap}{\varkappa}

\newcommand{\flux}{J}

\newcommand{\mbf}[1]{\mathbf{#1}}

\newcommand{\trm}[1]{\textrm{#1}}
\newcommand{\tsf}[1]{\textsf{#1}}

\newcommand{\be}{\begin{equation}}
\newcommand{\ee}{\end{equation}}
\newcommand{\bi}{\begin{itemize}}
\newcommand{\ei}{\end{itemize}}
\newcommand{\bea}{\begin{eqnarray}}
\newcommand{\eea}{\end{eqnarray}}
\newcommand{\tr}{\text{tr}\,}

\newcommand{\Ecr}{E_{\textrm{cr}}}

\newcommand{\nn}{\nonumber}
\newcommand{\sba}{\bar{s}}
\newcommand{\Fdual}{^{\ast}\!F}
\newcommand{\fdual}{^{\ast}\!f}

\newcommand{\figref}[1]{Fig. \ref{#1}}

\newcommand{\eqnref}[1]{Eq. (\ref{#1})}

\newcommand{\ket}[1]{|#1\rangle}
\newcommand{\proj}[2]{\langle #1 | #2 \rangle}

\newcommand{\braket}[3]{\langle #1 |#2 | #3 \rangle}

\newcommand{\vc}[1]{\mbox{\textbf{#1}}}

\newcommand{\ssvc}[1]{\mbox{\scriptsize\textbf{#1}}}
\renewcommand{\vec}[1]{\mathbf{#1}}

\newcommand{\smat}{\mathsf{S}}
\newcommand{\tmat}{\mathsf{T}}

\def\lambdabar{\protect\@lambdabar}
\def\@lambdabar{%
\relax \bgroup
\def\@tempa{\hbox{\raise.73\ht0
\hbox to0pt{\kern.2\wd0\vrule width.7\wd0
height.1pt depth.1pt\hss}\box0}}%
\mathchoice{\setbox0\hbox{$\displaystyle\lambda$}\@tempa}%
{\setbox0\hbox{$\textstyle\lambda$}\@tempa}%
{\setbox0\hbox{$\scriptstyle\lambda$}\@tempa}%
{\setbox0\hbox{$\scriptscriptstyle\lambda$}\@tempa}%
\egroup }

\begin{document}

\title{Coherent enhancement of QED cross-sections in electromagnetic backgrounds}

\author{T.~Heinzl}
\affiliation{Centre for Mathematical Sciences, Plymouth University, Plymouth, PL4 8AA, United 
Kingdom}
\author{B.~King}
\email{b.king@plymouth.ac.uk}
\affiliation{Centre for Mathematical Sciences, Plymouth University, Plymouth, PL4 8AA, United 
Kingdom}
\author{D.~Liu}
\affiliation{Centre for Mathematical Sciences, Plymouth University, Plymouth, PL4 8AA, United 
Kingdom}

\begin{abstract}
We introduce form factors that relate the amplitude of a QED process in vacuum to its corresponding background-field process. The latter is characterised by a reduced S-matrix element where one or more photon field operators are replaced by classical background fields. In the associated Feynman diagram, external photon lines are supplanted with lines representing the c-number field. This modifies the cross section by factors proportional to powers of the Fourier amplitude of the classical field (and its complex conjugate). We demonstrate this explicitly by comparing different reaction channels of low-energy photon-photon scattering in a classical background. We find that background field cross sections typically undergo coherent enhancement and for some reaction channels display a more favourable scaling with centre-of-mass energy compared to the vacuum process. 
Similar coherent enhancement may be found for leading-order pair annihilation to one photon, but this competes with kinematic suppression. This suppression can be minimised by using an x-ray free electron laser as the classical background.
\end{abstract}

\maketitle

\section{Context}
Some electromagnetic fields demonstrate a high degree of coherence. Examples include optical and x-ray free electron lasers, as well as slowly-varying fields such as those present in the vicinity of magnetars \cite{harding06}, in beam-beam collisions \cite{Blankenbecler:1988te} and in some oriented crystals \cite{Wistisen:2017pgr}. The effect of such fields on fundamental quantum electrodynamical (QED) processes can be calculated by describing them as classical background fields. As an immediate consequence, the background reduces the number of Poincar\`e symmetries of the vacuum and hence the number of conserved quantities \cite{Heinzl:2017zsr}. In particular, for realistic (non-constant) background fields, translation invariance will be (partially) broken so that four-momentum is in general not conserved. This allows processes that are kinematically forbidden in vacuum, such as pair annihilation to a single photon, to proceed in a background, provided kinematic constraints are met \cite{ilderton11b,Tang:2019ffe}. However, this may also imply that when processes already occurring in vacuum are `assisted' by a classical background, the assisted process has a more restricted kinematics than the vacuum process. On the one hand, an electromagnetic background can increase the probability of the assisted process when the background intensity is changed. On the other hand, the structure of the background (e.g. plane wave, focussed pulse, multiple pulses) may hinder the assisted process by reducing the outgoing phase space. In addition, the polarisation fraction of scattered particles can be altered when, in assisted processes, real photons in the vacuum process are replaced with an interaction with a background field of fixed polarisation. Therefore, QED processes in `low intensity' electromagnetic backgrounds, for which higher-order interactions with the background can be safely neglected, may yield interesting and potentially useful results.\\

X-ray free electron lasers (XFELs) such as the LCLS \cite{osti_1029479}, the EU.XFEL \cite{Abela:77248}, the SCSS at SACLA \cite{SACLA-yabuuchi2019experimental,SACLA-inoue2019generation}, FLASHForward \cite{ASCHIKHIN2016175}, SwissFEL \cite{etde_21413776,Schietinger:2024wsw} as well as planned XFELs such as the UK XFEL \cite{Dunning:2024nev} and SHINE \cite{Zhao:2024rig} have been suggested as tools to investigate QED due to their narrow bandwidth, high repetition rate and much higher energy than optical lasers. The potential of measuring assisted Compton scattering of x-ray photons in high-intensity \cite{Seipt:2013hda,Seipt:2015rda} and low-intensity \cite{Ahmadiniaz:2022mcy} optical backgrounds has been investigated. There are many suggestions for how to measure vacuum birefringence and real photon-photon scattering using an XFEL probe and an optical laser background \cite{dipiazza06,king10b,Karbstein:2015xra,Schlenvoigt:2016jrd,King:2018wtn,Karbstein:2022uwf,Ahmadiniaz:2022nrv,Macleod:2024jxl}. So far, experiments employing coherent fields have bounded the real photon scattering cross-section at eV \cite{Moulin:1999hwj}, keV \cite{Inada:2014srv,Yamaji:2016xws} and MeV \cite{Watt:2024brh} centre of mass energies, while others, such as BIREF@HIBEF \cite{Ahmadiniaz:2024xob} have been planned to improve upon these bounds. The current tightest bounds on the QED cross-section are provided by laser-cavity experiments such as PVLAS \cite{Ejlli:2020yhk} and BMV \cite{Agil:2021fiq}, where a quasi-constant field, coherent over macroscopic length scales, is employed. (Photon scattering at GeV energies has already been measured in Coulomb fields \cite{Jarlskog:1973aui,Akhmadaliev:1998zz,ATLAS:2017fur,CMS:2018erd,ATLAS:2019azn}, see also \cite{Ilderton:2024ufp}.)\\

In light of these developments aiming for tests of QED in laser fields, we revisit the link between QED in vacuum and QED in electromagnetic (EM) backgrounds. In particular, we want to explore how the standard QED process may be obtained from scattering in an EM background. Focussing on low-energy photon-photon scattering, we show that the cross-section for the field-assisted process with a background field interaction in the in and out state, (i) is enhanced by the intensity of the coherent background, (ii) has a different scaling with centre of mass (CM) energy, and (iii) produces a signal that is emitted in a different direction compared to the vacuum process. We argue that the modified scaling can be understood through an \emph{intensity form factor} in analogy to the charge density form factor of electron scattering. The intensity form factor takes a simple form in a plane wave background, but is modified by beam parameters when the background is focussed. We calculate the cross section for low-intensity assisted pair annihilation and find the same coherence factor as in photon-photon scattering. This coherence factor can be understood as a manifestation of \emph{bosonic enhancement}. Finally, example enhancement factors are calculated for colliding an electron beam with the EU.XFEL beam.
\newline

The paper is organised as follows. In Sec. II the form factor concept is developed for photon scattering off a classical wave, and in Sec. III the role of coherence in the enhancement of probabilities is explained. In Sec. IV-VII, the cross-sections for the $2\to2$, $1\to2$, $1\to1$ and $0\to1$ channels, respectively, are calculated. In Sec. VIII, the concept of form factors is applied to two-photon pair annihilation in a classical wave, and in Sec. IX the paper is concluded.

\section{Form Factors}

Scattering cross sections are at the heart of quantum field theory and particle physics. They measure the probability for elementary processes to happen and demarcate the point where theory and experiment have to meet: experimentalists indirectly measure the cross sections in the lab and compare with the theoretical predictions which typically have the general (differential) form,
\bea
  d\sigma = \frac{|\mathfrak{M}|^2}{J} d\Pi \; .
\eea
The invariant amplitude, $\mathfrak{M}$, is entirely determined by the underlying field theory, i.e.\ its Lagrangian and the parameters (masses, couplings) therein. The invariant phase space measure, $d\Pi$, encodes the kinematics of the process including energy-momentum conservation (if satisfied). To compare different experiments measuring the same process, one divides by the incident flux, $J$. Thus, there is a hierarchy in universality, beginning `on top' with the amplitude $\mathfrak{M}$ (defined by the fundamental interaction), followed by the phase space measure (defined by the process under consideration) and finally the flux (defined by the experiment).   \\

The connection between the invariant amplitude $\mathfrak{M}$ and the underlying field theory is provided by the S-matrix, defined through the Dyson series, $\smat = T \exp(-i \int dt \, H_I (t))$. Here, $T$ denotes time-ordering, and $H_I$ is the interaction Hamiltonian in the interaction picture. To remove trivial forward scattering it is customary to define the T-matrix via $\smat = 1 + i\tmat$. An experiment in essence measures scattering probabilities, hence (mod-squared) S-matrix elements between final and initials states,
\bea
  \braket{f}{\smat}{i} \equiv S_{fi} = \delta_{fi} + i \, T_{fi} \; .
\eea
At this point, there are two cases to consider. If all final and initial particle states correspond to plane waves with well-defined momenta ($p_f$ and $p_i$, respectively), one can factor out a momentum conserving delta function, which yields the standard invariant amplitude $\mathfrak{M}$ via
\bea \label{eqn:M.FI}
   S_{fi} = \delta_{fi} + (2\pi)^4 \, \delta (p_i - p_f) \, i \, \mathfrak{M}_{fi} \; .
\eea
We refer to this as a vacuum process. If, on the other hand, one scatters off a classical background field, say $\phi(x)$, its Fourier transform, $\tilde{\phi}(p)$, defines an extended momentum \emph{distribution} such that (\ref{eqn:M.FI}) gets replaced by 
\bea
  S_{fi} = \delta_{fi} + \tilde{\phi} (p_i - p_f) \, i \, \mathfrak{M}'_{fi} \; .
\eea
Initial and final momenta are now those of the incoming and outgoing \emph{particles} involved. This is consistent with Feynman rule (8) (for external fields) stated in \cite[Ch.~77]{landau4}.\\

A well known textbook example \cite{Halzen:1984} is given by electron-muon scattering, $e + \mu \to e' + \mu'$, where the scattering amplitude has the typical form involving two current matrix elements,
\bea
  T_{fi} &\sim& (2\pi)^4 \delta (p_i - p_f) \, \langle e' |j_\mu(0)| e \rangle \, \frac{1}{q^2} \, \langle \mu' |J^{\mu} (0)| \mu \rangle \nn \\
  & =& (2\pi)^4 \delta (p_i - p_f) \, \mathfrak{M}_{fi} \; .
\eea
Here $j_\mu$ and $J^\mu$ represent the electromagnetic currents of  electron and muon, respectively, both evaluated at the origin, $x=0$, while $q$ denotes the momentum transfer. Momentum conservation holds, $p_i = p_f$, with $p_i = p + k$ and $p_f = p' + k'$, where $p$ and $k$ are the initial muon and electron momenta respectively, with primes denoting the final momenta. \\

Consider now a classical field configuration, $A_\mu$, probed by an electron. This amounts to replacing the muon matrix element by a classical current, $ \langle \mu' |J^{\mu} (q)| \mu \rangle \to J^\mu (q) = -q^2 \tilde{A}^\mu (q)$, where $\tilde{A}^\mu (q)$ is the Fourier transform of the field. The scattering amplitude thus becomes
\bea \label{eqn:T-e-A}
  T_{fi} &\sim& \braket{e'}{j^\mu (0)}{e} \tilde{A}_\mu(q) \nn \\
  &=& - \braket{e'}{j^\mu (0)}{e} J_\mu(q)/q^2 \equiv \mathfrak{M}^\mu_{fi} \, \tilde{A}_\mu (q) \; .
\eea
In the static case, $\tilde{A}^\mu(q) = (2\pi) \delta^{\mu 0} \delta (q^0) \widetilde{\phi}(\vec{q})$, with $\widetilde{\phi}(\vec{q}) = J^0(\vec{q})/\vec{q}^2$, such that the amplitude takes on the Rutherford form, 
\bea
  T_{fi} &\sim& 2\pi \, \delta(q^0) \braket{e'}{j^0(0)}{e} J^0(\vec{q})/\vec{q}^2 \nn \\ 
  &\equiv&  2\pi \delta(q^0) \, \mathfrak{M}_{fi}  \, J^0(\vec{q})/\vec{q}^2 \; .
\eea
We see that the scattering amplitude factorises into the product of kinematic factors (the energy conserving delta function and the Fourier transform of the background field or current) times the charge density matrix element $\mathfrak{M}_{fi} = \braket{e'}{j^0}{e}$. For a point source, $J^0(\vec{q}) = 1$, so in general $J^0(\vec{q})$ measures the deviation from a point charge. For this reason, it is called a \emph{form factor}. Historically, form factors have been employed to characterise and measure the charge distribution of (static) nuclei \cite{Hofstadter:1956qs}. \\

This is not the end of the story, however. Form factors are useful beyond describing charge distributions: increasing the energy of the electron probes, they become sensitive to the substructure of nucleons given by quarks and gluons. The hadronic current matrix element, $J^\mu$, may then be parametrised in terms of scalar form factors or structure functions characterising the quark and gluon (`parton') distributions in hadrons and mesons. Put differently, they parametrise our theoretical ignorance, but can be determined experimentally. \\

The same logic is employed when the (renormalised) QED vertex is parametrised in terms of Dirac and Pauli form factors, $F_1(q)$ and $F_2(q)$, respectively, where $F_1(0) = 1$ (charge renormalisation condition), while $F_2$ determines (the QED part of) the anomalous magnetic moment of the electron. In this case, the form factors parametrise unknown higher-order contributions in QED, which (in principle, with enough effort) can be calculated. The current knowledge extends to five loops in QED perturbation theory, which corresponds to thousands of Feynman diagrams \cite{Aoyama:2019ryr}. \\

In this paper we are interested in the fundamental process of photon-photon (or light-by-light) scattering. Its occurrence, which is classically forbidden, is one of the earliest predictions of QED. To leading order in the electromagnetic coupling, there are four photons involved: $\gamma_1 + \gamma_2 \to \gamma_3 + \gamma_4$. The associated low-energy cross section was first determined by Heisenberg and his students \cite{euler35,heisenberg_euler36}; at the same time Landau's school analysed its high-energy behaviour \cite{akhiezer36,akhiezer37}. The general cross section, valid for all energies, was 
first calculated in the early 1950s \cite{karplus50,karplus51} and
is more elaborate due to the proliferation of Lorentz and helicity indices in the (mod-squared) amplitude. Interestingly, this theoretical prediction has not yet been confirmed for two reasons. For centre-of-mass energies of the order of MeVs (the gamma regime), the cross section $\sigma$ is sizeable, but photon fluxes $J$ are small. This renders photon counting rates, $N \sim J \sigma$, challenging to measure \cite{Watt:2024brh}. In the optical regime, where fluxes are large, the cross section itself is exceedingly small, which again leads to counting rates below the measuring threshold.\\

At this point, one tries to invoke the `help' of modern high-intensity lasers, which, having large field strengths can be well-modelled in terms of coherent states with high occupation numbers corresponding to a large photon flux. These coherent states can in turn be represented as external electromagnetic fields represented by classical c-numbers.  These are assumed to be under full experimental control, hence may be manipulated at will. Furthermore, as discussed above, the coupling to classical fields yields substantial simplifications for the S-matrix calculations of the amplitude $\mathfrak{M}$: some of the quantum mechanical matrix elements become trivial. The price to pay for the simplifications is a slight loss of universality as some matrix elements (or transition amplitudes) become inaccessible. \\

In this paper, we want to generalise the form factor concept to interactions with a source-less electromagnetic background, such as a plane wave or focussed laser pulse. Most of the paper will be centred on photon-photon scattering at low energy; later we will show how the concept can be applied to leading-order pair-annihilation. The interaction for low-energy photon-photon scattering can be described by a point-like four-photon vertex encoded in the Lagrangian (density)
\bea
  \mathcal{L}_\mathrm{eff} = c_{1} \mathcal{S}^{2} + c_{2} \mathcal{P}^{2} + c_{3} \mathcal{S}\mathcal{P} \; . \label{eqn:Leff1}
\eea
The first two terms arise in the leading-order weak-field expansion of the Heisenberg-Euler Lagrangian and the final term is a beyond-the-standard-model parity-violating contribution. (The field invariants are 
$\mathcal{S} = -F_{\mu\nu}F^{\mu\nu}/4$ and $\mathcal{P} = -\Fdual_{\mu\nu}F^{\mu\nu}/4$ where $F$ and $\Fdual$ are the field tensor and its dual \cite{itzykson80}.) The $c_i$ are low-energy constants: the first two of which, $c_1$ and $c_2$, can be expressed in terms of QED parameters but can also receive contributions from new physics such as Born-Infeld electrodynamics \cite{Davila:2013wba,Fouche:2016qqj} or axion-like-particles \cite{Ahlers:2007qf}. 

The Lagrangian (\ref{eqn:Leff1}) defines an $S$-matrix, $S = 1 + iT$, through its Dyson series. For the purposes of this paper, we are only interested in the leading-order contribution,
\bea \label{eqn:S-matrix}
  S^{(1)} \equiv i \int d^4 x \, \mathcal{L}_\mathrm{eff} (x) \; ,
\eea
which basically coincides with the leading order effective action, hence a single four-photon vertex. From this one can infer scattering amplitudes as usual by evaluating the corresponding T-matrix elements. For instance, the scattering amplitude for low-energy light-by-light scattering takes on the standard form \cite{Akhiezer:1965,detollis65,detollis71,landau4,itzykson80},
\bea \label{eqn:T-gamma-gamma}
  T_{fi}^{2 \to 2} = (2\pi)^4 \delta(k_1 + k_2 - k_3 - k_4) \, \mathfrak{M}^{2 \to 2}_{fi} \; , 
\eea
with the invariant amplitude given by
\bea
  \mathfrak{M}^{2 \to 2}_{fi} &\sim& \braket{\gamma_3 ,\gamma_4 }{\mathcal{L}_\mathrm{eff} (0)}{\gamma_1 , \gamma_2 } \nn \\
  &=& c_1 \braket{\gamma_3 ,\gamma_4 }{\mathcal{S}^{2} (0)}{\gamma_1 , \gamma_2 } + \ldots \; .
\eea
where $\mathcal{L}_{\trm{eff}}$ and $\mathcal{S}^{2}$ are evaluated at the origin, $x=0$. Working out the matrix elements with the help of Wick's theorem is lengthy and can be found in the literature \cite{karplus50,karplus51}. It is clear, though, that the Lorentz structure of the amplitude is given by \cite{Akhiezer:1965}
\bea \label{eqn:M-gamma-gamma}
  \mathfrak{M}^{2 \to 2}_{fi} &\sim& I_{\mu\nu\rho\sigma} \, \langle \gamma_3 ,\gamma_4 | A^\mu A^\nu A^\rho A^\sigma | \gamma_1 , \gamma_2 \rangle  \nn \\ 
  &\sim&  I_{\mu\nu\rho\sigma} (\varepsilon_1^\mu \varepsilon_2^\nu \varepsilon_3^{\ast\,\rho} \varepsilon_4^{\ast\,\sigma} + \ldots) \; ,
\eea
where the field operators $A$ are taken at the origin, $x=0$. The $\varepsilon_i$ are the polarisation vectors of the external photons, and we have suppressed all momentum arguments and helicity labels associated with the asymptotic states.  The fourth-rank tensor $I_{\mu\nu\rho\sigma}$ is constructed from all incoming and outgoing photon momenta which obey 4-momentum conservation according to (\ref{eqn:T-gamma-gamma}). \\

\begin{figure}[h!]
    \centering
    \includegraphics[width=\linewidth]{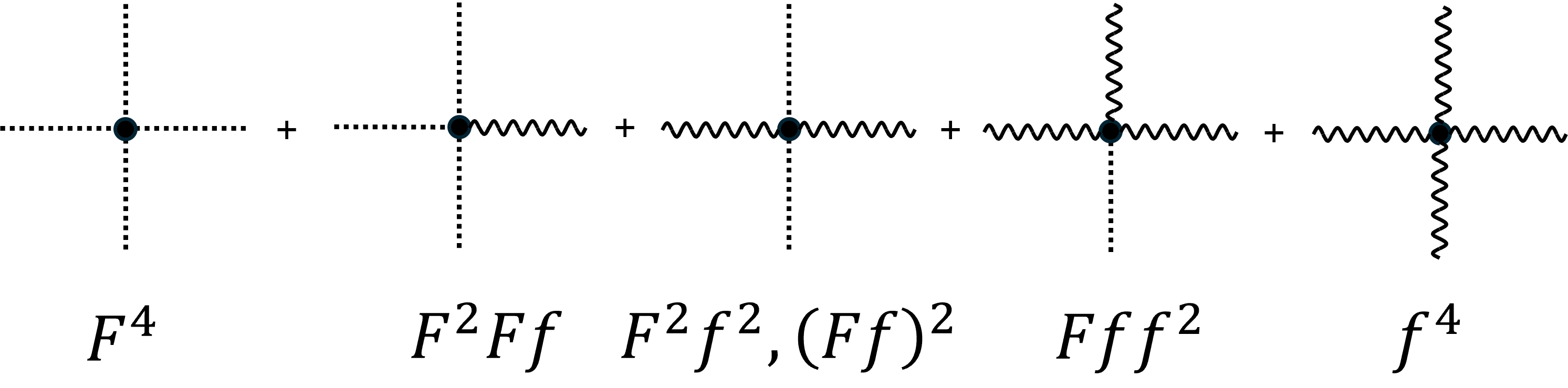}
    \caption{Feynman graphs showing all vertices contributing to $S^{(1)}$ (modulo permutations of legs), with zero to four external photons (wavy lines). Dotted lines denote the background.}
    \label{fig:Vertices}
\end{figure}

By analogy with scattering off a classical charge distribution, we want to consider the scattering of photons off a classical intensity distribution as provided, for instance, by a high-power laser. To this end we split all fields into a classical background, $A$, and a quantum fluctuation, $a$, through the replacements  $A \to A + a$ and, correspondingly, $F \to F + f$. Since the Lagrangian $\sim (F+f)^4$, the leading-order S-matrix (\ref{eqn:S-matrix}) turns into
\bea
  S^{(1)} \sim \int d^4 x \, \mathcal{L}_\mathrm{eff} [F+f] \; .
\eea
Focussing on the $c_1$ term for brevity, we have the expansion
\bea
  S^{(1)} \sim \int d^4 x \, &&\left[F^4 + 4 F^2 \, F\cdot f + 2 F^2 f^2 +4 (F\cdot f)^2 \right. \nn \\ 
  && \left. + 4 Ff \, f^2 + f^4\right] \;, 
\eea
where only the quantum field $f$ is operator valued. The $c_2$ and $c_3$ terms have analogous representations. In any case, this decomposition induces new vertices with any number of (quantised) photon lines ranging from zero to four, rendering the theory more intricate. Fig.~\ref{fig:Vertices} shows the associated vertex Feynman graphs. 
Each of these vertices has a number of matrix elements between photon states, with the number of external photon states no longer restricted to four, as they may be replaced by a background field. For instance, one can consider a transition from a one-photon to a one-photon state $(1 \to 1)$. Performing a Fock expansion of the photon field, $f$, all matrix elements may be evaluated using Wick's theorem. The resulting Wick contractions yield Feynman diagrams with one photon in both the initial and final states,
\bea
  S_{fi}^{1 \to 1} = \includegraphics[scale=0.39,valign=c]{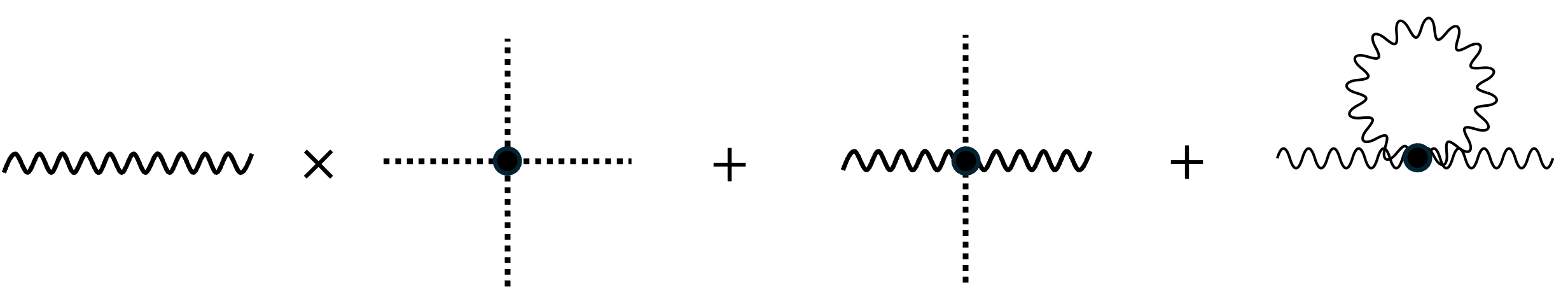} \nn \\
\eea
The first diagram is disconnected and does not describe photon scattering, while the second diagram does represent scattering off the background. The last diagram is a tadpole that may be eliminated by normal-ordering the interaction so that Wick self-contractions are avoided from the outset. If kept, the tadpole vanishes in dimensional regularisation \cite{Halter:1993kj} (but see \cite{Gies:2016yaa,Karbstein:2017gsb}). Furthermore, the first and the last diagram only have diagonal matrix elements in momentum space, hence contribute only to forward scattering, i.e. the $\delta_{fi}$ term in $S_{fi}$. The associated scattering amplitude is thus given by the middle diagram only,
\bea
  T_{fi}^{1 \to 1} \equiv \includegraphics[scale=0.45,valign=c]{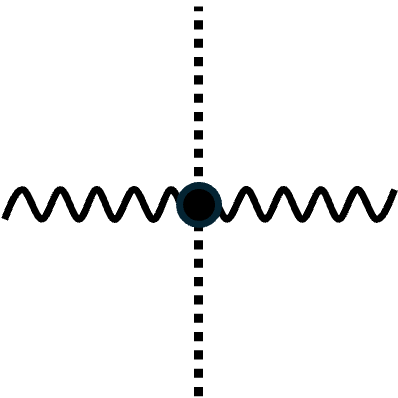} \; .
\eea

\begin{figure}[h!!]
\centering
\includegraphics[width=8.4cm]{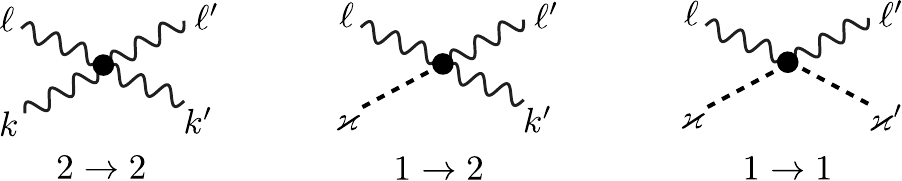}
\caption{Channels for low-energy photon-photon scattering in the presence of a classical background. Wavy lines represent photons and dashed lines the classical field with $\vkap$ denoting the momentum from a Fourier mode of the field.} \label{fig:FeynmanEff}
\end{figure}

The discussion above shows that there are several channels of interest for the scattering of $n \to m$ photons from the quantum fluctuation field, $a$, which are kinematically supported. The most relevant of these channels are displayed in \figref{fig:FeynmanEff} together with external momentum assignments. Note that the dashed (background) lines do \emph{not} end in the usual `crosses' (representing localised sources) as the background is source-free and solves the vacuum wave equation.
\newline 

In what follows we discuss these channels one by one.

\subsection{$1 \to 2$ photon scattering}

If we consider the $1 \to 2$ channel, $\gamma \to \gamma'_1 + \gamma'_2$ (\figref{fig:FeynmanEff}, central panel), the only connected T-matrix element is
\bea \label{eqn:T-gamma-laser1t2}
  &&T_{fi}^{1 \to 2} \sim \int d^4 x \, c_1 \, \braket{\gamma'_1\gamma'_2}{\mathfrak{s}(x) f^{\rho\sigma} (x)}{\gamma} \, F_{\rho\sigma}(x) + \ldots \nonumber \\
  &&\equiv c_1 \braket{\gamma'_1\gamma'_2}{\mathfrak{s}(0) f^{\rho\sigma} (0)}{\gamma} \,  \tilde{F}_{\rho\sigma} (q) + \ldots \equiv \mathfrak{M}^{1\to2}_{\rho\sigma} \,  \chi^{\rho\sigma} (q) \;, \nn \\
  \label{eqn:T12}
\eea
where we have defined the fluctuation invariant, $\mathfrak{s} \equiv f^{\mu\nu}f_{\mu\nu}$ to save space. For this process, $q = k'+\ell' - k$ is the momentum transfer of an incoming probe photon with momentum $k$ scattering off the classical field to produce two outgoing photons of momentum $k'$ and $\ell'$ respectively. The terms in ellipsis denote other contributions from the Lagrangian linear in $F$. In analogy with \eqnref{eqn:T-e-A}, describing $e$-$\mu$ scattering, $\tilde{F}_{\rho\sigma} (q)$ is the Fourier transform of the background field,
\bea \label{eqn:chi1t2}
  \tilde{F}_{\rho\sigma} (q)  &=& \int d^4 x \, e^{i q \cdot x} \, F_{\rho\sigma}(x) \equiv  \chi_{\rho\sigma} (q) \; .
\eea
and so $\chi_{\rho\sigma}$ acts as a \emph{field-dependent form factor}. Let us assume that the tensor structure of the classical background is constant and can be factored out. This is the case if the background comprises multiple weakly-focussed fields or plane waves. Let us chose, for example, a background configuration characterised by a light-like momentum $\vkap^\mu = \omega_\vkap n^\mu$, $\omega_\vkap \equiv \vkap^0$,
\bea 
  F^{\mu\nu}(x) = F_0 \, g(x) \, \epsilon^{\mu\nu} \;, \quad \epsilon^{\mu\nu} = n^{\mu}\epsilon^{\nu} - n^{\nu}\epsilon^{\mu} \;, 
\label{eqn:FmunuSimp}
\eea
with field amplitude $F_0$ and dimensionless profile $g(x)$. This leads to the form factor
\bea
  \chi_{\mu\nu}(q) = F_0 \, \tilde{g}(q) \, \epsilon_{\mu\nu} \equiv \chi^{(1)} (q) \, 
  \epsilon_{\mu\nu} \; ,
\eea 
which manifestly scales like the field amplitude $F_0$. The scattering amplitude (\ref{eqn:T12}) thus becomes
\bea
  T_{fi}^{1 \to 2} = F_0 \, \tilde{g}(q) \, \epsilon^{\mu\nu} \mathfrak{M}_{\mu\nu}^{1 \to 2} \; .
\eea
For a plane wave background such as (\ref{eqn:FmunuSimp}), with profile $g=g(\vkap \cdot x)$ depending only on the single light-front variable $\vkap\cdot x$, the classical and quantum external lines in the Feynman diagrams (dashed and wavy lines in Fig.~\ref{fig:FeynmanEff}, respectively) have the same form but a different normalisation. As a result, one can obtain the $1 \to 2$ amplitude, $\mathfrak{M}_{\mu\nu}^{1 \to 2}$, from the $2 \to 2$ amplitude, $\mathfrak{M}^{2\to2}$, by a simple prescription: replace one of the incoming photons (i.e., its momentum and polarisation) in $\mathfrak{M}^{2\to2}$ with the classical field (momentum $\vkap$ and polarisation $\epsilon$), hence
\bea
  \epsilon^{\mu\nu} \mathfrak{M}_{\mu\nu}^{1 \to 2} = 
  \sqrt{\frac{2V}{\omega_{\vkap}}} \left[\mathfrak{M}^{2\to2}\right]_{1\to2} \; ,
  \label{eqn:M.12.M.22}
\eea
where the last expression represents the replacement rule. This results in an alternative representation for the $1 \to 2$ scattering amplitude,
\bea 
  T_{fi}^{1 \to 2} = \sqrt{\frac{2V}{\omega_{\vkap}}}\chi^{(1)}(q)\left[\mathfrak{M}^{2\to2}\right]_{1\to2} \; .
\eea

\subsection{$1 \to 1$ photon scattering}

In the same vein, the connected $1\to 1$ amplitude can be written as
\bea \label{eqn:T-gamma-laser}
  T_{fi}^{1 \to 1} &\sim& \int d^4 x \, c_1 \, \braket{\gamma'}{f^{\mu\nu}(x)f^{\rho\sigma} (x)}{\gamma} \, F_{\mu\nu}(x) F_{\rho\sigma}(x) + \ldots \nonumber \\
  &\equiv& c_{1}\braket{\gamma'}{f^{\mu\nu}(0)f^{\rho\sigma} (0)}{\gamma} \,  \chi_{\mu\nu\sigma\rho} (q) +\ldots \nn \\[5pt]
  &\equiv& \mathfrak{M}^{1\to1}_{\mu\nu\rho\sigma} \,  \chi^{\mu\nu\sigma\rho} (q),
\eea
with momentum transfer $q = \ell' - \ell$.  Re-employing the analogy with (\ref{eqn:T-e-A}), the tensor $\chi_{\mu\nu\rho\sigma}$ is the Fourier transform of the product of background fields (hence a convolution in momentum space),
\bea \label{eqn:Post}
  \chi_{\mu\nu\rho\sigma} (q) &=& \int d^4 x \, \mbox{e}^{i q \cdot x} \, F_{\mu\nu}(x) F_{\rho\sigma}(x)  \nn \\
  &\equiv& \left[\tilde{F}_{\mu\nu} \ast \tilde{F}_{\rho\sigma}\right](q) \; .
\eea
This may be interpreted as a field dependent incarnation of Post's general constituent tensor \cite{Post:1962} for nonlinear electrodynamics. Being \emph{quadratic} in the classical field, it describes the \emph{intensity} distribution of the background, and thus can be viewed as an \emph{intensity form factor}, in contrast to the $1 \to 2$ form-factor (\ref{eqn:chi1t2}) which is linear in the background. It should be emphasised that the amplitude (\ref{eqn:T-gamma-laser}) does not involve 4-momentum conservation. However, the Post tensor $\chi$ acts as a `momentum filter': it needs to have support at the photon momentum transfer, $q = \ell'-\ell$. In other words, the background momentum distribution, $\chi(q)$, provides a whole spectrum of momenta that can be transferred between incoming and outgoing photons. If $F$ is close to the form of a plane wave, as for a weakly-focussed laser pulse, then $\chi(q)$ only has support around $q\approx 0$. In this case, assuming the form (\ref{eqn:FmunuSimp}), Post's tensor (\ref{eqn:Post}) becomes 
\bea
  \chi_{\mu\nu\rho\sigma}(q) = F_0^2 \, 
  \left[\tilde{g}(q) \ast \tilde{g}(q)\right] \, \epsilon_{\mu\nu} \epsilon_{\rho\sigma}
  \equiv \chi^{(2)}(q) \,
  \epsilon_{\mu\nu} \epsilon_{\rho\sigma} \;, \nn \\
\eea
and scales as the field amplitude squared. The scattering amplitude (\ref{eqn:T-gamma-laser}) thus turns into 
\bea 
  T_{fi}^{1\to1} = \frac{2V}{\omega_{\vkap}}\chi^{(2)}(q)\left[\mathfrak{M}^{2\to2}\right]_{1\to1}, \label{eqn:Tfi1to1simp}
\eea
where now the prescription on the $2\to2$ matrix element means to replace one incoming and one outgoing photon momentum and polarisation with the classical field momentum $\vkap$ and its polarisation $\epsilon$, i.e.,
\bea
    \frac{2V}{\omega_{\vkap}} \left[\mathfrak{M}^{2\to2}\right]_{1\to1} = \epsilon^{\mu\nu} \epsilon^{\rho\sigma} \mathfrak{M}^{1\to1}_{\mu\nu\rho\sigma} \; .
\eea

\subsection{$1 \to 0$ photon scattering}

Finally, one can also define the $1\to 0$ process of photon absorption in the classical background. Because of the larger number of classical fields involved, the tensor structure is more complicated,
\bea \label{eqn:T-gamma-laser1t0}
  T_{fi}^{1 \to 0} &\sim& c_1  \braket{0}{f^{\mu\nu}(x)}{\gamma} \chi^{(3)}_{\mu\nu}(q)  \nn \\
  && + \braket{0}{\fdual^{\mu\nu}(x)}{\gamma}\left[c_{2}^{\ast}\!\chi^{(3)}_{\mu\nu}(q)+c_{3}\chi^{(3)}_{\mu\nu}(q)\right],
\eea
where $q=-\ell$ only and $\chi^{(3)}$ is a nonlinear form factor depending on the third power of the field:
\bea
  \chi^{(3)}_{\mu\nu} (q) &=& \int d^4 x \, \mbox{e}^{i q \cdot x} \, F^{\rho\sigma}(x)F_{\mu\nu}(x) F_{\rho\sigma}(x) \nn \\
  &\equiv& \left[\tilde{F}^{\rho\sigma} \ast \tilde{F}_{\mu\nu} \ast \tilde{F}_{\rho\sigma}\right] (q) \; .
\eea
and $^{\ast}\!\chi^{(3)}$ is the same with one contracted field tensor replaced with its dual. If $F$ approximates the field strength of a plane wave, then $F^{\rho\sigma}F_{\rho\sigma} \approx 0$, which implies $\chi^{(3)}_{\mu\nu}(q) \approx 0$. Hence, the form of the classical background can strongly affect the kinematics, to the point that there is almost complete kinematic suppression of the $1 \to 0$ process in a weakly-focussed laser pulse. However, if the background comprises two plane waves, $F\to F_{1} +F_{2}$ with $F_{1}\cdot F_{2} \neq 0$ then with a suitable choice of wave vectors, the $1\to 0$ process need not be suppressed.

\section{Coherent Enhancement}

The results of the previous section may be rephrased in terms of photon coherent states. In the present context, they were introduced by Kibble \cite{Kibble:1965zza} and quite recently employed in \cite{Copinger:2024pai}. The coherent states are eigenstates of the photon annihilation operator, $a_{\ssvc{k},s}$, (with momentum $\vc{k}$, helicity $s$) and hence of the positive frequency part of the photon field operator,
\bea
  \hat{\vc{A}}(x) &=& \hat{\mathbf{A}}^{(+)}(x) + \text{h.c.}  \nonumber \\
  &=& \sum_{\ssvc{k},s}  \frac{1}{\sqrt{2\omega_{\ssvc{k}}V}}
  \, a_{\ssvc{k},s} \mathbf{\epsilon}_{\ssvc{k},s} \, e^{-i k \cdot x} + \text{h.c.} 
  \; .
\eea
Here we have assumed a finite quantisation volume, $V$, implying discrete momenta, $k = (\omega_{\ssvc{k}}, \vc{k})$. We have also adopted radiation gauge, $A^0=0$, $\boldsymbol{\nabla} \cdot \vc{A} = 0$. Denoting the coherent state by $|z\rangle$, we thus have 
\bea
  a_{\ssvc{k},s} \, |z\rangle &=& z_{\ssvc{k},s} \, |z\rangle \; ,
\eea
where the eigenvalue represents the c-number Fourier mode, $z_{\ssvc{k},s}$, of the `classical field', described by the classical potential $\vc{A} = \langle z | \hat{\vc{A}} | z \rangle$, and replaces the Fock operator $a_{\ssvc{k},s}$. The coherent states $|z \rangle$ may be generated by acting with a unitary displacement operator on the vacuum, $| z \rangle = D_z | 0 \rangle$, where
\bea
  D_z  = \exp \sum_{\ssvc{k},s}  \left\{ z_{\ssvc{k},s} a_{\ssvc{k},s}^\dagger - z_{\ssvc{k},s}^* a_{\ssvc{k},s} \right\} \; .
\eea
The crucial displacement property is encoded in the shift relation,
\bea \label{eqn:DISP}
  D_z^\dagger a_{\ssvc{k},s} D_z = a_{\ssvc{k},s} + z_{\ssvc{k},s} \; .  
\eea
As a result, the S-matrix elements in the presence of a classical background field strength, $F=dA$, can be expressed as \cite{Kibble:1965zza}
\bea \label{eqn:S.SPLIT}
  \langle f | S [\hat{f} + F] | i \rangle = \langle f |D_z^\dagger S[\hat{f}] D_z | i \rangle \equiv  
  \langle f, z |S[\hat{f}]| z, i \rangle \; , \nonumber \\
\eea
where $\hat{f} = d\hat{A}$ is the field strength operator. We thus note that the decomposition into background and quantum fluctuation performed in the previous section is equivalent to the determination of matrix elements between coherent states. For this to make sense it is important to note that the coherent states $|z\rangle$ depend on the background via the background Fourier amplitudes $z_{\ssvc{k},s}$ which appear in the displacement operator (\ref{eqn:DISP}). Let us assume the background is a monochromatic plane wave, with momentum $\kappa$ and polarisation $\sigma$ which we will simply refer to as `the laser'. This implies that
\bea
  z_{\ssvc{k},s} = z_{\boldsymbol{\kappa}} \delta_{\ssvc{k}\boldsymbol{\kappa}} \delta_{s\sigma} \; .
\eea
This obviously means that the coherent state will be annihilated by all modes $a_{\ssvc{k},s}$ apart from the laser mode, $a_{\boldsymbol{\kappa},\sigma}$. As an example, for a monochromatic laser field with field strength $F^{\mu\nu} = F_0 \cos (\kappa \cdot x) (\epsilon^\mu n^\nu - \epsilon^\nu n^\mu)$, $n^\mu = \kappa^\mu/\omega_{\boldsymbol{\kappa}}$, the Fourier amplitude (of the potential) is $z_{\boldsymbol{\kappa}}= i F_0 \sqrt{ V/2\omega_{\boldsymbol{\kappa}}}$.

To avoid clutter we simplify the situation considerably by getting rid of all indices and momentum arguments: we consider a toy model consisting of two modes, laser and non-laser, with commuting annihilators $a$ and $b$, respectively.  Coherent states, $|z \rangle = D_z |0 \rangle$, are defined with respect to the laser mode $a$ only, hence via the displacement operator
\bea \label{eqn:DISP2}
  D_z = \exp (z a^\dagger - z^* a) \; .
\eea
We thus have $b |z \rangle = 0$ and $a |z \rangle = z |z \rangle$, with the eigenvalue $z$ representing the classical field amplitude. At any given order in perturbation theory, the quantised S-matrix will be a function of both modes, $S = S[a, a^\dagger; b, b^\dagger]$. We thus consider the scattering amplitudes
\bea \label{eqn:S.SPLIT2}
  S_{fi} = \langle f, z | S[a, a^\dagger; b, b^\dagger] | z, i \rangle \; ,
\eea
where both final and initial states are assumed to be a product of $b$-number states and laser coherent states,
\bea
  | f, z \rangle = (b^\dagger)^n |z \rangle \; , \quad | i, z \rangle = (b^\dagger)^m |z \rangle \; .
\eea
They thus contain a finite number of $b$-photons on top of a coherent state of laser photons (the background). Using the displacement operator (\ref{eqn:DISP2}), we can rewrite (\ref{eqn:S.SPLIT2}) by moving the background dependence into the $S$ matrix as in (\ref{eqn:S.SPLIT}),
\bea
  S_{fi} = \langle f| S[a + z, a^\dagger + z^*; b, b^\dagger] | i \rangle \; .
\eea
The terms in the S-matrix will be monomials in $a$, $b$ and their hermitian conjugates. A typical term may thus be written as a current-current interaction of the form $S = j(a, a^\dagger) J(b, b^\dagger)$ where both $a$ and $b$ currents are assumed to be normal-ordered to eliminate tadpoles (self-contractions). We thus have the S-matrix element
\bea
  S_{fi} &=& \langle f, z | j(a, a^\dagger) J(b, b^\dagger) | z, i \rangle \nonumber \\
  &=& \langle z |j(a, a^\dagger) | z \rangle \langle f |J(b, b^\dagger) |i \rangle \equiv j(z, z^*) \, J_{fi} \; ,
\eea
which is the product of a classical background current, $j(z, z^*)$ and a transition current. For the probability one thus has the mod-square,
\bea
  \tsf{P}_{z,fi} = |j(z, z^*)|^2 |J_{fi}|^2 \; .
\eea
This result may be interpreted as a \textit{coherent enhancement} of the $b$-mode probability, hence the probability without background, $\tsf{P}_{0,fi} = |J_{fi}|^2$. In the enhancement factor, $|j(z, z^*)|^2$, the background current $j(z, z^*)$ typically contains terms like $|z|^2 = \bar{n} \equiv \langle z | a^\dagger a | z \rangle$, the mean photon number in the coherent state. The situation is very much akin to the bosonic enhancement for absorption transitions in highly occupied number states, $|n \rangle = (a^\dagger)^n|0\rangle$, 
\bea
  \tilde{S}_{fi} = \langle f, n-1 | a J(b, b^\dagger) | n, i \rangle = \sqrt{n} \, J_{fi} \; ,
\eea
which implies an absorption probability enhanced by a factor of $n$ \cite{Gottfried-Yan:2004}.

As a concrete example for coherent enhancement let us compare the background free $2\to2$ process to the background $1\to1$ process in the two-mode toy model. For the leading-order S-matrix we adopt the particle number conserving expression $\hat{S}_{ab} = a^\dagger a b^\dagger b$. If, for the $2\to2$ process, we assume a transition between just number states $|1,1\rangle := a^\dagger b^\dagger |0\rangle$, the transition amplitude becomes\footnote{A slight bosonic enhancement is observed if one considers a transition between doubly occupied states, such as $|2\rangle := (a^\dagger)^2 |0\rangle$: $S_{fi} = \langle 2 | \hat{S}_{aa} | 2 \rangle = \langle 2 | a^\dagger a^\dagger a a | 2 \rangle = 2$, and analogously for $b$-modes.}
\bea
   S_{fi} = \langle 1,1 | \hat{S}_{ab} | 1,1 \rangle = \langle 1 | a^\dagger a | 1 \rangle \langle 1 | b^\dagger b | 1 \rangle = 1 \; . 
\eea
This should be compared with a transition between $b$-number, $a$-coherent states, $|z,1\rangle = b^\dagger |z\rangle$, corresponding to the $1\to1$ process for $b$-modes in the presence of a background, 
\bea
  S_{fi} = \langle z,1 | \hat{S}_{ab} | z, 1 \rangle = \langle z| a^\dagger a | z \rangle \langle 1| b^\dagger b | 1 \rangle = |z|^2 \; , 
\eea
hence a coherent enhancement by a factor $|z|^2 = \bar{n}$. This \emph{quantum} coherence is different from \emph{classical} coherence: in reality, the overall coherent enhancement also depends on the `intrinsic' coherence of the background field. For example, suppose the classical background is approximately monochromatic and can be written as a sum of $N$ monochromatic plane waves that are slightly shifted with respect to each other by a phase, $\phi_{j}$, i.e.,
\bea
F^{\mu\nu}(x) = F_{0} \eps^{\mu\nu} \sum_{j=1}^{N} \cos \left(\vkap \cdot x + \phi_{j}\right), \label{eqn:N.SUP}
\eea
where $\eps^{\mu\nu}\eps_{\mu\nu}=0$ as in (\ref{eqn:FmunuSimp}). For $N=1$, this implies a Fourier amplitude $z_\vkap = iF_0 \sqrt{V/2\omega_\vkap} \exp(-i\phi_1)$, with mod square equal to the mean photon number,
\bea
  |z_\vkap|^2 = F_0^2V/2\omega_\vkap \equiv \bar{n}_\vkap \; . 
\eea
For $N >1$, this gets modified according to
\bea
  |z_\vkap|^2 = \bar{n}_\vkap \sum_{ij}^N \exp i \Delta_{ij} \equiv \bar{n}_\vkap \, C_0 \; , \;\; \Delta_{ij} \equiv \phi_i - \phi_j \; , \nn \\
\eea
where the sum of relative phases, $C_0 = \sum \exp \Delta_{ij}$, may be evaluated as
\bea
  C_0 = N + 2 \sum_{i < j} \cos \Delta_{ij} \approx \left\{ 
  \begin{array}{ll}
      N^2 \; , &  \Delta_{ij} \ll 1 \\
      N \; , & \Delta_{ij} \;\; \text{random} \,. 
      \label{eqn:C0}
  \end{array}
  \right. 
  \nn \\
\eea
It thus measures the coherence of the $N$ superimposed waves (\ref{eqn:N.SUP}). Looking at 
the intensity form factor, $\chi_{\mu\nu\sigma\rho} (q) = \eps_{\mu\nu}\eps_{\sigma\rho}\chi(q)$, with
\bea
  \chi(q) \sim 
  C_{2}\delta(2\vkap + q)+2C_{0}\delta(q)+C_{-2}\delta(-2\vkap + q) \;,
  \nn \\
\eea
we see the coherence factor $C_0$ reappearing for zero momentum transfer, $q=0$. Since the incoming and outgoing photons are on-shell, this channel is the most relevant. According to (\ref{eqn:C0}), when the relative phases, $\Delta_{ij}$, are much smaller than unity, then $C_{0} \approx N^{2}$ and we obtain a further enhancement from the coherence of the $N$ waves. On the other hand, when the relative phases are randomly distributed, the field is incoherent, and $C_{0}$ only scales like $N$. Therefore, the degree of coherence of the electromagnetic background during the interaction with the photon is crucial in providing an enhancement of the photon-photon process compared to scattering from an incoherent source of photons. (Coherence also plays an important role in the emitted radiation field of electrons in laser fields, see e.g. \cite{Gelfer:2023mdz}.)

\section{Cross-section of 2-to-2 photon scattering}

All our results for photon scattering in a classical field will be compared to the same baseline process, namely light-by-light scattering ($2 \to 2$) in vacuum. Let us briefly review how to calculate the associated cross section at low energies \cite{Akhiezer:1965,itzykson80,Landau:1987gn}. We use the effective Lagrangian from \eqnref{eqn:Leff1} and the momentum labels from \figref{fig:FeynmanEff} (left-most panel), so $k + \ell \to k' + \ell'$. Following \cite{itzykson80,Davila:2013wba}, the interaction is rewritten using the identity
\be \label{eqn:TR.F}
\left(\trm{tr}\,F\Fdual\right)^{2} = 4\, \tr F^{4} - 2\left(\tr F^{2}\right)^{2} \; ,
\ee
where the field strength $F$ is viewed as a 4-by-4 matrix. Decomposing this field into a sum,
\be \label{eqn:F.DECOMP}
F = F_{\ell}+F_{k} + F_{\ell'}+F_{k'},
\ee
with each incoming photon field of the form
\[
F^{\mu\nu}_{\ell} = \frac{1}{\sqrt{2\ell^{0}V}}\,\left[\ell^{\mu}\eps_{\ell}^{\nu} - \ell^{\nu}\eps_{\ell}^{\mu}\right] \mbox{e}^{-i \ell \cdot x},
\]
(and the complex conjugate for the outgoing fields) allows to bypass the full machinery of Wick's theorem: one just has to insert the decomposition (\ref{eqn:F.DECOMP}) into the 4-photon vertex employing (\ref{eqn:TR.F} and keep only terms containing all four fields from (\ref{eqn:F.DECOMP}). The invariant amplitude, $\mathfrak{{M}}$, thus becomes a sum of traces which may be evaluated by summing over polarisations such that the unpolarised result takes on the form
\bea \label{eqn:M.2.2}
  \left|\mathfrak{M}^{2\to2}_{fi}\right|^{2} 
  &=& \frac{278\alpha^2}{2025} 
  \left[  (\ell \cdot k)^{4} + (\ell\cdot \ell')^{4} + (\ell\cdot k')^{4} 
  \right] \; ,\nn \\
\eea
assuming standard QED values for the parameters $c_{j}$. 
(Explicit expressions for the polarised result have been presented elsewhere in the literature \cite{karplus50,karplus51,detollis65,Davila:2013wba}.) The result can be written in terms of Mandelstam variables, 
\bea
  s &=& (\ell+k)^2 = 2 \ell \cdot k \; , \nn \\
  t &=& (\ell-\ell')^2 = -2 \ell \cdot \ell' \equiv q^2 
  \nn \; , \\
  u &=& (\ell-k')^2 = -2 \ell \cdot k' \; . \label{eqn:MANDELSTAM}
\eea
The invariant amplitude (\ref{eqn:M.2.2}) thus takes on the compact form
\bea 
\left|\mathfrak{M}^{2\to2}_{fi}\right|^{2} 
&=& \frac{139\alpha^{2}}{16200}\left(s^{4}+t^{4}+u^{4}\right). \label{eqn:M2mandel}
\eea
The unpolarised differential probability of photon-photon scattering is 
\bea
\tsf{P}^{2\to 2} = \frac{1}{4}\sum_{\trm{pol.}}V^{2} \int \frac{d^{3}\pmb{\ell'}\,d^{3}\mbf{k'}}{(2\pi)^{6}}\,|T^{2\to2}_{fi}|^{2} \; , \label{eqn:P2to2aa}
\eea
or, expressed in terms of the scattering amplitude:
\[
T_{fi}^{2\to2} = i(2\pi)^{4}\delta^{(4)}(q) \, \mathfrak{M}_{fi}^{2\to2}
\]
with momentum transfer $q=k'+\ell'-k-\ell$. The delta function takes care of the $d^{3}\mbf{k'}$ integral in \eqnref{eqn:P2to2aa}. Writing $d^{3}\pmb{\ell'} = \omega'^{2} d\omega' d\Omega'$ with $\omega' = |\pmb{\ell'}|$, we use the remaining delta function in energy to perform the  integral in $\omega'$, hence
\bea
\tsf{P}^{2\to 2} = \frac{1}{4}\frac{1}{(2\pi)^{2}}\sum_{\trm{pol.}}V^{3}T \int\, d\Omega'~(\omega'_{\ast})^{2}~|\mathfrak{M}^{2\to2}_{fi}|^{2}, \label{eqn:P2to2bb}
\eea
where $k' = (k'^{0}_{\ast},\mbf{k'}_{\ast})$, and asterisks denote quantities fixed by momentum conservation, $\mbf{k'}_{\ast} = \pmb{\ell}+\mbf{k}-\pmb{\ell'}$,  $k'^{0}_{\ast} = |\mbf{k'}_{\ast}|$ and $\omega'_{\ast} = \ell^{0}+k^{0}-k'^{0}_{\ast}$. $V$ and $T$ are the usual divergent volumetric factors arising in $\left[\delta^{(4)}(q)\right]^{2} = \delta^{(4)}(q) VT/(2\pi)^4$.\\

It is useful to evaluate Lorentz invariants in the centre-of-momentum (CM) frame, following Karplus and Neuman \cite{karplus51}: We thus employ the momenta
\[
\ell = \omega(1,0,0,1);\quad k = \omega(1,0,0,-1);
\]
\[
\ell' = \omega(1,\sin\theta,0,\cos\theta);\quad k' = \omega(1,-\sin\theta,0,-\cos\theta),
\]
and choose linear polarisation basis states $\eps^{(1)} = (0,0,1,0)$ (for all photons) and
\[
  \eps_{\ell}^{(2)} = (0,1,0,0); \quad \eps_{k}^{(2)} = (0,1,0,0); 
\]
\bea
  \eps_{\ell'}^{(2)} = (0,-\cos\theta,0,\sin\theta); \quad \eps_{k'}^{(2)} = (0,\cos\theta,0,-\sin\theta). \nn \\ \label{eqn:poldef}
\eea
The $2\to2$ cross-section, $\sigma_{2\to 2}$ is obtained via
\bea
  \frac{d\sigma^{2\to 2}}{d\Omega'} = \frac{V}{T} \frac{1}{v_{\trm{rel.}}}\frac{d\tsf{P}^{2\to 2}}{d\Omega'}, \label{eqn:dsigma2to2}
\eea
where $v_{\trm{rel.}}$ is the relative velocity of the colliding particles \cite{cannoni17}. Noting that $\mathfrak{M}_{fi}^{2\to2} \sim 1/V^4$, all volumetric factors are seen to cancel. The relative velocity of particles with lightlike momenta $\ell_{1}$ and $\ell_{2}$ is $v_{\trm{rel.}} = \ell_{1}\cdot \ell_{2}/\omega_{1}\omega_{2}$, which in the CM frame takes the simple form $v_{\trm{rel.}}=2$ \cite{Coleman:2018mew}. Defining the classical electron radius, $r_{e} = \alpha/m$, we find
\bea 
  \frac{d\sigma^{2\to 2}}{d\Omega'} &=& \frac{m^{2}r_{e}^{2}}{\alpha^{2}}\,m^{6}\frac{2(c_{1}-c_{2})^{2}+(c_{1}+c_{2})^{2}+2c_{3}^{2}}{64 \pi^{2}} \nn \\
  && \qquad \times \left(3+ \cos^{2}\theta_{3}\right)^{2}\left(\frac{\omega}{m}\right)^{6}, \label{eqn:dsigmaA1}
\eea
which agrees with results in the literature \cite{Rebhan:2017zdx}.
Inserting the leading-order QED values for the low-energy constants $c_{1}=8\alpha^{2}/45m^{4}$ and $c_{2}=14\alpha^{2}/45m^{4}$, $c_{3}=0$ \cite{karplus50} yields\footnote{We keep $c_{3}=0$ for the remainder of the paper.}
\bea 
\frac{d\sigma^{2\to 2}}{d\Omega'} = \frac{139\,\alpha^{2}r_{e}^{2}}{32400 \pi^{2}}\left(3+ \cos^{2} \theta_{3}\right)^{2}\left(\frac{\omega}{m}\right)^{6},
\eea
agreeing with e.g. \cite{karplus51,detollis65}. (The \emph{differential} cross-section includes a factor $2$ to take into account identical photons in the out state.) In terms of Mandelstam variables, one finds the compact and symmetric expression
\bea
\frac{d\sigma^{2\to 2}}{d\Omega'} = \frac{\alpha^{2}r_{e}^{2}}{2\pi^2\omega^2}\frac{139}{45^{2}}(s^4+t^4+u^4)  \label{eqn:dsMandelstam}
\eea
As $s+t+u=0$, one may rewrite $s^4+t^4+u^4 = 2(s^2t^2+s^2u^2+t^2u^2)$ to show that \eqnref{eqn:dsMandelstam} is equivalent to the alternative expression given by De Tollis \cite{detollis65}. When integrated over scattering angles, the total cross-section is\footnote{A factor of $1/2$ has been multiplied onto the \emph{total} cross-section to allow for identical outgoing particles \cite{detollis65}}:
\bea 
  \sigma^{2\to 2} = \frac{973\,\alpha^{2}r_{e}^{2}}{10125 \pi}\left(\frac{\omega}{m}\right)^{6}. \label{eqn:sig2to2a}
\eea
We note from \eqnref{eqn:dsigmaA1} that the full cross-section for $2\to2$ scattering involves both the difference and the sum of the low-energy constants. In comparison, the probability for vacuum birefringence, a feature of the $1\to1$ process under current investigation \cite{Ahmadiniaz:2024xob}, only depends on the \emph{difference} of the low-energy constants \cite{Toll:1952rq}. This is a first hint that the cross-section for photon scattering in a classical field will differ in general from the $2\to 2$ (standard QED) cross-section.
\newline 

In what follows, we repeat the previous calculation for photon scattering in a classical field.\\

\section{Cross-section of 1-to-2 photon scattering}

Let us begin with the probability of the $1 \to 2$ process, sometimes referred to as leading-order `photon splitting' in a classical field. We focus in this section on \emph{polarised} photons because one incoming photon has been replaced by the classical field, which has fixed polarisation. The probability for $1 \to 2$ scattering of photons off a classical field with polarisation $\epsilon$ is then
\bea
  \tsf{P}^{1\to 2}_{\epsilon} = V^{2} \int \frac{d^{3}\pmb{\ell'}\,d^{3}\mbf{k'}}{(2\pi)^{6}}\,|T_{fi,\epsilon}^{1\to2}|^{2}. 
  \label{eqn:P1to2a}
\eea
Since we have replaced an incoming photon by a classical field, there is no overall momentum-conserving delta-function that can be used to perform the outgoing momentum integrals as in the $2\to2$ case. However, we can view the classical field as a wave packet with a momentum distribution given by its Fourier transform. Then, for each Fourier mode with momentum $\vkap$, we have four-momentum conservation of the form
\[
  \ell + \vkap = \ell' + k'.
\]
Using the product form \eqnref{eqn:T12} for the scattering amplitude, writing the form factor as
\bea
  \chi_{\sigma\rho}(q) = \int d^{4}k~ \chi_{\sigma\rho}(\vkap) \delta(\vkap-q), \label{eqn:chi-trick-1}
\eea
and casting the $d^{3}\ell'$ integral into spherical polar form, the $d^{3}\mbf{k'}$ and $d|\pmb{\ell}'|$ integrals can be performed to give
\bea
  \tsf{P}^{1\to 2}_{\epsilon} &=& \frac{V^{2}}{(2\pi)^{6}} \int d^{4}\vkap\,\chi_{\sigma\rho}(\vkap)\chi_{\sigma'\rho'}^{\ast}(\vkap) \nn \\
  && \times \int\, d\Omega'~\omega'^{2}_{\ast}~ \mathfrak{M}_{fi,\epsilon}^{\sigma\rho}\left[\mathfrak{M}_{fi,\epsilon}^{\sigma'\rho'}\right]^{\dagger},
  \label{eqn:P1to2b}
\eea
where we have used that $k' = (k'^{0}_{\ast}, \mbf{k'}_{\ast})$, with components $\mbf{k'}_{\ast} = \pmb{\ell} + \pmb{\vkap}-\pmb{\ell'}$, $k'^{0}_{\ast} = |\mbf{k'}_{\ast}|$ and $\omega'_{\ast} = \ell^{0}+\vkap^{0}-k'^{0}_{\ast}$. The result (\ref{eqn:P1to2b}) can be related to the $2\to2$ process as follows: we recall from \eqnref{eqn:chi1t2} that the form factor is the Fourier transform of the field,
\be
  \chi_{\rho\sigma} = \tilde{F}_{\rho\sigma}  =   -i\left(\vkap_{\rho} \tilde{A}_{\sigma}-\vkap_{\sigma}\tilde{A}_{\rho}\right) \; ,
\ee
where the background potential, $\tilde{A}$, is a solution of the vacuum wave equation, $\vkap^2 \tilde{A} = 0$, and hence has the form
\bea
  \tilde{A}^{\rho}(\vkap) = 2\pi \delta(\vkap^{2})\,\bar{A}^{\rho}(\vkap)\; .
  \label{eqn:Abardef1}
\eea
For a single wave with linear polarisation along $\epsilon^\rho_{\vkap}$, one has $\bar{A}^{\rho}(\vkap)=\epsilon^\rho_{\vkap}\bar{A}(\vkap)$.
The Fourier amplitude $\bar{A}$ specifies the type of background (e.g. plane wave, focussed pulse, etc.). For this choice of background and form factor, the probability (\ref{eqn:P1to2b}) turns into
\bea
  \tsf{P}^{1\to 2}_{\epsilon} &=& \frac{1}{(2\pi)^{6}}\frac{1}{T} \int d^{4}\vkap\,(2\vkap^{0})|2\pi \delta\!\left(\vkap^2\right)\! \bar{A}(\vkap)|^{2}~ \tsf{P}^{2\to2}_{\epsilon}(\vkap) \;,\nn \\ 
  \label{eqn:P1to2d} 
\eea
where $\tsf{P}^{2\to 2}_{\epsilon}$ is the polarised version of \eqnref{eqn:P2to2bb} with one incoming photon replaced by the classical field of fixed polarisation $\epsilon$. Evaluating the squared delta-function using $\delta^2(\vkap^{0}\pm |\pmb{\vkap}|) = T\delta(\vkap^{0}\pm |\pmb{\vkap}|)/2\pi$, we obtain
\bea
  \tsf{P}^{1\to 2}_{\epsilon} &=& \int \frac{d^{3}\pmb{\vkap}}{(2\pi)^{3} 2 \omega_\vkap} \, 2|\bar{A}(\vkap)|^2 \, \tsf{P}^{2\to2}_{\epsilon}(\vkap) \; , 
\label{eqn:P1to2dd}
\eea
where $\vkap$ is on shell, $\vkap = (\omega_\vkap, \pmb{\vkap})$. We thus see that, for the $1\to2$ process, the Fourier amplitude of the classical potential acts like a `wave packet of photons', hence a `classical' photon momentum distribution (of prescribed polarisation), 
\be \label{eqn:WEIGHT}
  2|\bar{A}(\vkap)|^2 \equiv |z_\vkap|^2 \; , \quad \vkap^2 = 0
  \; .
\ee
It is thus clear how the classical background affects the probability, but does it modify the cross-section? Note first that it is not suitable to define a cross-section as in \ref{eqn:dsigma2to2} for the $2 \to 2$ process, since for the $1\to 2$ process there is no relative velocity (the kinematics are that of a decay). However, if we define the differential cross section as $d\sigma = dP/j$ where $j = JT/V$ denotes the incoming flux density and $d\tsf{P}$ the differential scattering probability, we can extend the wave packet analogy by writing
\be
  \frac{d\sigma^{1\to2}}{d\Omega'} = \frac{1}{\flux}\frac{V}{T} \frac{d\tsf{P}^{1\to2}}{d\Omega'} \;.
\ee
(For the $2\to2$ case, the photon wave functions have already been normalised to one particle per unit volume so $\flux = 1$.) The incoming number flux can be defined as the `zeroth moment' of the distribution function (\ref{eqn:WEIGHT}) or the normalisation integral,
\be
  J = \int \frac{d^{3}\pmb{\vkap}}{(2\pi)^{3} 2 \omega_\vkap} \, 2 |\bar{A}(\vkap)|^2 \; .
  \label{eqn:fluxA}
\ee
This is consistent with the 4-momentum $\Pi^{\mu}$ carried by the classical field (assuming Lorenz gauge),
\bea
  \Pi^{\mu} = \int \frac{d^{3}\pmb{\vkap}}{(2\pi)^{3} 2 \omega_\vkap} \, 2|\bar{A}(\vkap)|^2 \, k^\mu \; .
  \label{eqn:fluxB}
\eea
where the same momentum distribution function appears as in (\ref{eqn:P1to2dd}).

To show the feasibility of this approach, we calculate the cross-section for a laser pulse modelled as a monochromatic plane wave propagating down the $z$-axis \cite{Waters:2017tgl} with Fourier amplitude
\bea 
  \bar{A}(\vkap) &=& -i(2\pi)^{3}F_{0}\theta(\vkap^{3})\delta\left(\pmb{\vkap}^{\perp}\right) \nn \\ 
  && \qquad\times \left[\delta(\omega_{\vkap}-\vkap^{0})+\delta(\omega_{\vkap}+\vkap^{0})\right]. \label{eqn:Amono}
\eea
This corresponds to a dimensionless flux $\flux = F_{0}^{2} \mathsf{A} T/2 \omega_{\vkap}$, where $\mathsf{A}$ is a cross-sectional area that arises from $\delta(\pmb{\vkap}^{\perp})^2 = \delta(\pmb{\vkap}^{\perp}) \mathsf{A}/(2\pi)^2$. Inserting (\ref{eqn:Amono}) into (\ref{eqn:P1to2dd}), the flux factor exactly cancels, which results in
\bea 
  \frac{d\sigma^{1\to2}}{d\Omega'} = \left[\frac{d\sigma^{2\to2}}{d\Omega'}\right]_{1\to 2} \; . \label{eqn:dsig1}
\eea
The prescription on the right amounts to replacing one of the incoming photon momenta and polarisations of the $2\to 2$ process by the fixed polarisation and wave vector of the monochromatic wave, $(k_\mathrm{in}, \varepsilon_\mathrm{in}) \to (\vkap, \epsilon_\vkap)$. It is the appearance of the same distribution function in (\ref{eqn:fluxA}) or (\ref{eqn:fluxB}) and in the $1\to 2$ cross-section through (\ref{eqn:P1to2dd}) which guarantees that the $1\to 2$ cross-section in a classical background is not significantly different from that of the vacuum process.

\section{Cross-section of 1-to-1 photon scattering}

\subsection{General results}

One major difference of the $1 \to 1$ process is the reduction in dimension of the outgoing phase space as can be seen from the probability,
\bea
  \tsf{P}^{1\to 1} = V \int \frac{d^{3}\pmb{\ell'}}{(2\pi)^{3}}\,|T_{fi}^{1\to1}|^{2}. \label{eqn:P1to1a}
\eea
Following the discussion in the introduction, see (\ref{eqn:T-gamma-laser}), the scattering amplitude, $T_{fi}^{1\to1}$, can be written as
\bea
  T_{fi}^{1\to1} &=& \chi^{\mu\nu\rho\sigma}(q) \,\mathfrak{M}^{1\to 1}_{\mu\nu\rho\sigma}
\eea
with intensity form factor 
\bea
  \chi^{\mu\nu\rho\sigma}(q) = \int d^{4}x\,\mbox{e}^{iq\cdot x}F^{\mu\nu}(x)F^{\rho\sigma}(x) \label{eqn:chiPARAX}
\eea
and invariant amplitude 
\bea
  \mathfrak{M}^{1\to 1}_{\mu\nu\rho\sigma} = \frac{c_1}{4} \langle \ell' , \epsilon' | f_{\mu\nu} (0) f_{\rho\sigma} (0)| \ell, \epsilon \rangle + \cdots \; ,
\eea
identifying $q = \ell' - \ell$ as the momentum transfer.
Splitting the the photon fields $f$ into positive and negative frequency parts and introducing the one-particle wave functions 
\bea
  f_{\mu\nu}^{\trm{in}} &=& \langle 0 | f_{\mu\nu} (0) | \ell, \epsilon  \rangle = -i(\ell_{\mu}\epsilon_{\nu} -\ell_{\nu}\epsilon_{\mu}) \;  , \\
  f_{\mu\nu}^{\trm{out}} &=& \langle \ell', \epsilon' | f_{\mu\nu} (0) |  0 \rangle = i(\ell'_{\mu}\epsilon'_{\nu} -\ell'_{\nu}\epsilon_{\mu}') \; ,
\eea
the invariant amplitude assumes the compact form
\bea 
 \mathfrak{M}^{1\to 1}_{\mu\nu\rho\sigma} = \frac{c_{1}}{4} \left(f^{\trm{in}}_{\mu\nu}f^{\trm{out}}_{\rho\sigma}+f^{\trm{out}}_{\mu\nu}f^{\trm{in}}_{\rho\sigma}\right) +  \cdots 
\eea
A weakly-focussed paraxial Gaussian pulse, for which the Infinite Rayleigh Length Approximation (IRLA) \cite{King:2012aw,King:2018wtn,Karbstein:2021ldz,Berezin:2024fxt} is applicable, has a (background) field strength of the form (\ref{eqn:FmunuSimp}), $F_{\mu\nu} = F_0 \, g(x) \epsilon_{\mu\nu}$ with profile function
\bea
   g(x) = e^{-\left[ \Delta_{1} x_1^2 + \Delta_{2} x_2^2 + \Delta_{-} (x^-)^2\right]/2 } \, \cos (\omega_{\vkap}x^{-})
   \; . \label{eqn:F-paraxial}
\eea
Note that we allow for $\Delta_1 \ne \Delta_2$, hence we do not assume azimuthal symmetry -- the beam may have an elliptic cross section, for instance. The pulse propagates in $z$ direction with its phase determined by the light-front coordinate $x^- = \vkap \cdot x/\omega_\vkap$. It is characterised by the beam waists, $w_{0,1}$ and $w_{0,2}$ in $1$ and $2$ directions, the pulse duration $\tau$ and the focussing parameters, 
\bea \label{eqn:DELTAS}
  \Delta_{1} = 2/w_{0,1}^2; \quad \Delta_{2} = 2/w_{0,2}^{2}; \quad  \Delta_{-} = 2/\tau^{2} \; . 
\eea
The co-ordinates have been chosen such that the three spatial axes align, respectively, with the electric, magnetic and propagation directions of the classical field. Hence, the intensity form factor in \eqnref{eqn:chiPARAX} can be written in the form (\eqnref{eqn:Tfi1to1simp}),
\bea 
  \chi^{\mu\nu\rho\sigma}(q) &=& \epsilon^{\mu\nu} \epsilon^{\rho\sigma} \chi^{(2)}(q) \; ,
\eea 
with scalar form factor
\bea 
  \chi^{(2)}(q) &=& \frac{(F_{0}\pi)^{2}}{2}\sqrt{\frac{\pi}{\Delta_{1}\Delta_{2}\Delta_{-}}} ~ \delta(q_{+})\, \nn \\ 
  && \times \exp\left[ -\frac{(q_{1})^{2}}{4\Delta_{1}}-\frac{(q_{2})^{2}}{4\Delta_{2}} 
  - \frac{(q_{-})^{2}}{4\Delta_{-}}\right] \; . \label{eqn:FF.2}
\eea
Here we have evaluated a sum over frequency components and kept only those terms which correspond to one absorption and one emission from the classical field. Note that the momentum transfer has vanishing plus component: $q_+ = q^-/2 = (q^0 - q^3)/2 = 0$. To interpret the result (\ref{eqn:FF.2}), compare with the case of a monochromatic plane-wave background for which all focussing parameters, $\Delta_{1,2,-}$, in (\ref{eqn:F-paraxial}) vanish. In this case, the intensity form-factor in a monochromatic plane wave, $\chi_{\tsf{mono}}$ assumes the simple form
\bea 
  \chi^{(2)}_{\tsf{mono}}(q) = (2\pi F_{0})^{2} \delta^{(4)}(q),
\eea
where, again, we have discarded terms corresponding to two absorptions or two emissions from the classical field. The effect of the background field profile can be clearly seen: in the focussed Gaussian case, the form factor supports a range of momenta transferred to the emitted photon, whereas in the monochromatic case, only elastic forward scattering $\ell'=\ell$ is allowed, which amounts to vanishing momentum transfer. (Scattering is not trivial, though, as typically there are changes in polarisation, see below.)
\newline

For the focussed background profile (\ref{eqn:F-paraxial}) the probability becomes a convolution integral,
\bea 
  \tsf{P}^{1\to 1} = V \int \frac{d^{3}\pmb{\ell'}}{(2\pi)^{3}} \, |\chi^{(2)}(\ell' - \ell)|^{2} \, Q(\ell'), \label{eqn:P.11.CONV}
\eea
where the second factor is
\bea 
  Q(\ell') &=& \frac{1}{\omega_{\vkap}^{2}} \left|\epsilon^{\mu\nu} \epsilon^{\rho\sigma} \, \mathfrak{M}_{\mu\nu\rho\sigma}^{1\to 1}\right|^{2} \equiv \frac{(2V)^2}{\omega_{\vkap}^2} \left| \left[\mathfrak{M}^{2\to 2} \right]_{1\to 1} \right|^{2} 
  \nn \\
\label{eqn:Qlp1}
\eea
To perform the integration over $\ell'$ in (\ref{eqn:P.11.CONV}), we change integration variables from Cartesian to light-front using
\bea
  \int d^{3}\pmb{\ell'}  &=& \int d^{4}\ell'~2\ell'^{0}\,\theta\left(\ell'^{0}\right) \delta\left(\ell'\cdot\ell'\right)  \nn \\
  &=& 4\int d\ell'_{+}d\ell'_{-}d^{2}\ell'_{\perp} \ell'^{0}\,\theta\left(\ell'^{0}\right) \delta\left(4\ell'_{+}\ell'_{-}-\ell'_{\perp}\cdot\ell'_{\perp}\right)  \nn \\
  &=& \int \frac{d\ell'_{+}d^{2}\ell'_{\perp}}{\ell'_{+}} \ell'^{0}_{\ast}\,\theta\left(\ell'^{0}_{\ast}\right) 
\eea
where $\ell^{\prime\,0}=\ell'_{+}+\ell'_{-}$, and in $\ell_{\ast}^{\prime\,0}$, the delta function has been evaluated with $\ell'_{-} \to \ell'_{\perp}\cdot\ell'_{\perp}/4\ell'_{+}$. The $d\ell'_{+}$ integral is then performed using the $\delta(q_{+})$ in $\chi^{(2)}(q)$, which results in
\bea 
  \tsf{P}^{1\to 1} &=& \frac{\pi^{2}}{2^{5}} 
  \frac{V F_{0}^{4}}{\Delta_{1} \Delta_{2} \Delta_{-}} 
  \, \int d\ell'_{1} \, d\ell'_{2} \, \frac{\ell^{\prime 0}}{\ell_{+}} \, \delta_{+}(0) \, Q(\ell') \nn \\ 
  && \times \exp \left[ -\frac{q_1^2}{2\Delta_{1}} - \frac{q_2^2}{2\Delta_{2}} - \frac{1}{2\Delta_{-}} \left( \frac{\ell_{\perp}^{\prime 2}}{4\ell_{+}} - \ell_{-} \right)^{2} \right]. \nn \\
\eea
The delta function with vanishing argument is interpreted as
\bea
  \delta_{+}(0) = \frac{\lim_{\ell'_{+}\to 0}\delta(\ell'_{+})\lim_{\ell'_{-}\to 0}\delta(\ell'_{-})}{\lim_{\ell'_{-}\to 0}\delta(\ell'_{-})}=\frac{L}{2\pi}\frac{\ell^{0}}{2\ell_{+}}\nn \\
\eea
with a longitudinal length factor, $L$, so that the probability becomes
\bea 
  \tsf{P}^{1\to 1} &=& \frac{\pi}{2^{5}}\frac{V L F_{0}^{4} }{\Delta_{1} \Delta_{2} \Delta_{-}}\, \int d\ell'_{1}\,d\ell'_{2} \,\frac{\ell'^{0}\ell^{0}}{(2\ell_+)^{2}}\,Q(\ell') \nn \\ 
  && \times \exp\left[-\frac{(\ell'_{1})^{2}}{2\Delta_{1}}-\frac{(\ell'_{2})^{2}}{2\Delta_{2}}-\frac{1}{2\Delta_{-}}\left(\frac{(\ell'_{\perp})^{2}}{4\ell_{+}}-\ell_{-}\right)^{2}\right]. \nn \\
\eea
Writing the momentum transfer in transverse direction as $\vc{q}_\perp = (q_1, q_2) = (X\sqrt{\Delta_{1}}, Y\sqrt{\Delta_{2}})$, introduces new integration variables $X$, $Y$, through
\bea
  \ell_{1}'&=&\ell_{1} + X\sqrt{\Delta_{1}} \nn \\
  \ell_{2}'&=&\ell_{2} + Y\sqrt{\Delta_{2}},
\eea
and results in
\bea 
  \tsf{P}^{1\to 1} &=& \frac{\pi}{2^{5}}\frac{V L F_{0}^{4} }{\Delta_{-}\sqrt{\Delta_{1} \Delta_{2}}}\frac{1}{(2\ell_{+})^{2}} \nn \\
  && \times \int dX\,dY \exp\left[-\frac{X^{2}}{2}-\frac{Y^{2}}{2}\right] R(X,Y; \mathbf{\Delta})  \nn \\
  \label{eqn:P.11.FOCUSSED}
\eea
with abbreviations $\mathbf{\Delta} = (\Delta_1, \Delta_2)$ and
\bea 
  R(X,Y;\mathbf{\Delta}) &=&
  \mbox{e}^{ -\frac{1}{2\Delta_{-}}\left(\frac{\Delta_{1}X^{2} + \Delta_{2} Y^{2} + 2\sqrt{\Delta_{1}} \ell_{1}X + 2 \sqrt{\Delta_{2}} \ell_{2}Y}{4\ell_{+}} \right)^{2} } \nn \\ 
  && \times \ell'^{0}\ell^{0} Q\left(\ell_{1}+ X\sqrt{\Delta_{1}},\ell_{2}+Y\sqrt{\Delta_{2}}\right), \nn \\ \label{eqn:R2e}
\eea
The transverse momentum arguments of $Q = Q(\ell'_{1},\ell'_{2})$ have been written out explicitly. The results (\ref{eqn:P.11.FOCUSSED}) and (\ref{eqn:R2e}) describe the full scattering probability in the focussed background (\ref{eqn:F-paraxial}) without azimuthal symmetry. The expressions are somewhat involved as the background with all its parameters describes a fairly realistic configuration. To simplify things, we use our assumption that the focussing is \emph{weak}. We can then Taylor expand the function $R$ from (\ref{eqn:R2e}) in the transverse focussing parameters, $\Delta_{1}$ and $\Delta_{2}$. The zeroth-order term corresponds to $q = \ell' - \ell = 0$, i.e., the probability in a plane-wave background. To leading order we thus have
\bea 
  \tsf{P}^{1\to 1} &=& \frac{\pi^{2}}{2^{4}} 
  \frac{V L F_{0}^{4} }{\Delta_{-} \bar{\Delta}} ~ \frac{(\ell^{0})^{2}}{(2\ell_{+}^{2})} \Bigg\{ \! \left( 1-\frac{\Delta_{1} \ell_{1}^2 + \Delta_{2} \ell_{2}^2}{8\Delta_{-}\ell_{+}^2} \right) Q(\ell_{1},\ell_{2}) \nn \\
  && + \frac{\Delta_{1}}{2} \frac{\partial^{2}Q(\ell_{1}',\ell_{2})}{\partial \ell_{1}^{\prime\,2}} \Bigg|_{\ell_{1}' = \ell_{1}} \!
  + \frac{\Delta_{2}}{2}\frac{\partial^{2} Q(\ell_{1},\ell_{2}')}{\partial \ell_{2}^{\prime\,2}} \Bigg|_{\ell_{2}'=\ell_{2}} \Bigg\} \; , \nn \\
\label{eqn:P1to1foc4}
\eea
where, in the prefactor, $\bar{\Delta} \equiv \sqrt{\Delta_1 \Delta_2}$ denotes the geometric mean. To interpret the expression (\ref{eqn:P1to1foc4}), we compare with the $2\to2$ amplitude squared from (\ref{eqn:Qlp1}). We therefore replace the momenta and polarisations in the $2\to2$ process (see \figref{fig:FeynmanEff}, left-most panel) according to
\be
  k, k' \to \vkap; \quad \eps_{k}, \eps_{k'} \to \epsilon \; ,
\ee
which turns the Mandelstam variables (\ref{eqn:MANDELSTAM}) into 
\bea
  s &=& -u = 2 \ell \cdot \vkap \; \nn , \\ 
  t &=& -\mbf{q_{\perp}}^{2} = -\left(X^2 \Delta_1 + Y^2 \Delta_2\right)
\eea
This tells us that in a weakly-focussed background, the (squared) momentum transfer, $t = q^2$, is directly proportional to the focussing parameters $\Delta_{1,2}$, whereas in  a plane wave (no focussing), one has $t=0$ identically since $q = 0$. The momentum transfer $q$ thus measures the deviation from the plane wave case. Going back to (\ref{eqn:P1to1foc4}), the term proportional to $Q(\ell_{1},\ell_{2})$ contains focussing corrections which originate from the intensity form factor, $\chi^{(2)}$, but an amplitude that is evaluated as for free photons. The terms containing derivatives introduce focussing corrections to the amplitude itself, by adding momentum transfer originating from the focussing. However, for the case at hand, we find the derivative terms are zero and only the correction from the intensity form factor survives.

\subsection{Application}

\subsubsection{Linearly polarised background}

To compare our findings for the $1\to1$ probability to literature values, we consider the background to be linearly polarised in the $1$-direction, i.e., $\epsilon = \epsilon_{1}$ with $x \cdot \epsilon_{j} = -x^{j}$ for $j = 1,2$. We assume the incoming photon has helicity corresponding to vectors $\eps_{\pm}=(\eps_{1}\pm i\eps_{2})/\sqrt{2}$ with momentum dependent basis,
\bea
  \eps_{j} = \epsilon_{j} - \frac{\ell \cdot \epsilon_{j}}{\vkap\cdot \ell} \, \vkap \; , \quad \ell \cdot \eps_j = 0 
  \; .
\eea
Adopting these choices together with forward scattering, $\ell'=\ell$, we find the invariant amplitude (squared),
\bea 
  \left|\left[\mathfrak{M}^{2\to 2}\right]_{1\to 1}\right|^{2}  = \left(\frac{s^{2}(c_{1}\pm c_{2})}{8V^2\,\ell^{0}\omega_{\vkap}}\right)^2 \label{eqn:M1to1squared}
\eea
where the sum (difference) of low-energy constants corresponds to the case of helicity no-flip (flip). For the scattering probability (\ref{eqn:P1to1foc4}) we obtain
\bea 
  \tsf{P}^{1\to 1} &=& \frac{\pi^{2}}{2^{6}} 
  \frac{F_{0}^{4}}{\omega_{\vkap}^{2} A \Delta_{-} \bar{\Delta}} 
  \frac{s^{4}}{(4\ell_{+}\omega_{\vkap})^2} 
  \left(1-\frac{\Delta_{1}\ell_{1}^2 + \Delta_{2}\ell_{2}^2}{8\Delta_{-}\ell_{+}^2}\right) \nn \\
  && \qquad\times\begin{Bmatrix}
    (c_{1}+c_{2})^2 \\
    (c_{1}-c_{2})^2 
  \end{Bmatrix}
  \label{eqn:P1to1foc5}
\eea
where $A = V/L$ is the transverse area.  The lower (upper) expression in braces corresponds to the probability of helicity-flip (no-flip). Inserting Mandelstam $s = 2\vkap \cdot \ell = 4 \omega_{\vkap} \ell_{+}$, the physical meaning of the focussing parameters  from (\ref{eqn:DELTAS}) and the QED values for the low-energy constants, we end up with the probability 
\bea 
  \tsf{P}^{1\to 1} &=& \frac{\pi^{2}}{2^{8}} 
  \frac{F_{0}^{4} \tau^{2} w_{0}^{2}}{Am^{4}\omega_{\vkap}^{2}} \left(1-\frac{\tau^{2}(\pmb{\ell_{\perp}})^2}{2w_{0}^{2}(\ell^{-})^2}\right) \left(\frac{2\alpha^{2}s}{45m^{2}}\right)^2\begin{Bmatrix}
    11^2 \\
    3^2
\end{Bmatrix} . \nn \\
\eea
This can be recast into the form
\bea 
\tsf{P}^{1\to 1} &=& \frac{\pi w_{0}^{2}}{2A} \left(1-\frac{\tau^{2}(\pmb{\ell^{\perp}})^2}{2w_{0}^{2}(\ell^{-})^2}\right) \left(\frac{\alpha \eta\xi^{2}\Phi}{90\pi m}\sqrt{\frac{\pi}{2}}\right)^2\begin{Bmatrix}
    11^2 \\
    3^2
\end{Bmatrix}, \nn \\
\label{eqn:P.11.VB}
\eea
with the usual background intensity parameter, $\xi = eF_{0}/m\omega_{\vkap}$. We have also employed the energy parameter, $\eta$, in the form 
\bea
  \eta = \vkap\cdot \ell/m^2 = s/2m^2 = 2 \omega^2/m^2 \; , 
  \label{eqn:ETA}
\eea
where $\omega$ is the CM frequency, and $\Phi$ denoting the pulse phase duration. If we define the transverse area as $ A = \pi w_{0}^{2}/2$ (which follows from calculating the energy density), and take the limit of zero focussing, the probability (\ref{eqn:P.11.VB}) agrees with known literature values \cite{dinu14a,king16}.
\smallskip

With the probability established, let us turn to the $1\to1$ cross-section, which we define as
\bea
  \sigma^{1\to1} = \frac{1}{\flux}\frac{V}{T}\tsf{P}^{1\to1} 
  \; . \label{eqn:SIGMA.11.A}
\eea
For a weakly-focussed Gaussian beam, the incoming flux as obtained from (\ref{eqn:fluxB}) is 
\bea 
  \flux = \frac{F_{0}^{2}}{\omega_{\vkap}} \left( \frac{\pi^{3}}{\Delta_{-} \Delta_{1} \Delta_{2}} \right)^{1/2} = \frac{F_{0}^{2}w_{0}^{2}\tau}{\omega_{\vkap}}\left(\frac{\pi}{2}\right)^{3/2}, \label{eqn:fluxGauss1}
\eea
where we have used the focussing parameters (\ref{eqn:DELTAS}). Using this together with the probability (\ref{eqn:P.11.VB}), the cross-section (\ref{eqn:SIGMA.11.A}) becomes
\bea 
  \sigma^{1\to1} &=& \sqrt{\frac{2}{\pi}}\frac{\xi^{2}\Phi}{\alpha} \left(1-\frac{\Phi^2\pmb{\ell^{\perp}}^2}{2 \mu^{2}(\ell^{-})^2}\right) \nn \\
  && \qquad\times \left(\frac{2\alpha r_{e}}{45}\right)^{2}\left(\frac{\omega}{m}\right)^{4}\begin{Bmatrix} 11^{2} \\3^{2}\end{Bmatrix}\,, 
  \label{eqn:sig.11}
\eea
where $V=AT$ has been used, and $\mu=\omega_{\vkap}w_{0}$ is the transverse phase associated with focussing. The ratio of cross sections,
\bea 
  \frac{\sigma^{1\to 1}}{\sigma^{2\to 2}} \sim \frac{\xi^{2}\Phi}{\alpha\eta}\left(1-\frac{\Phi^2\pmb{\ell^{\perp}}^2}{2 \mu^{2}(\ell^{-})^2}\right), \label{eqn:sig1to1foc}
\eea 
may be called the `coherence factor' as it describes an enhancement of the scattering probability, in particular for intense fields with $\xi \gg 1$. The appearance of beam-dependent quantities, $\xi^{2}\Phi$, in the cross-section is a result of the $1\to 1$ interaction having the classical field associated with the \emph{out} state so that division by the incoming flux does not cancel all such beam factors. This is in contrast to the $1 \to 2$ process, where the classical field is associated with the \emph{in} state, and beam factors are exactly cancelled, see e.g. \eqnref{eqn:dsig1}. One may conclude that when the laser field is scattered \emph{into}, the cross-section is no longer universal and instead depends on parameters of the beam.
\newline

A striking difference between the $1\to 1$ and $2\to 2$ processes is the scaling of the cross-sections with CM energy: 
\bea
  \sigma^{1\to 1} &\sim& (\omega/m)^4 \sim \eta^2 \; , \nn \\
  \sigma^{2\to 2} &\sim& (\omega/m)^6 \sim \eta^3 \; .
  \label{eqn:sig.scaling}
\eea
Note this change in scaling does not arise from any flux factor; it stems from the fact that the kinematics of the two processes process are different. A similar situation arises for Delbr\"uck scattering, i.e.\ the scattering of a real photon on the Coulomb field of a nucleus of charge $Z$. In the low-energy limit, the 
total cross-section scales like $\eta^2$ \cite{detollis71},
\bea
  \sigma_{\trm{D}} = \frac{1}{3\pi}(Z^{2}\alpha)^{2}\left(\frac{\alpha r_{e}}{1152}\right)^2
  \left(\frac{\omega}{m}\right)^{4}
  \begin{Bmatrix} 73^{2} \\45^{2}\end{Bmatrix}~, \label{eqn:sigDel}
\eea
where the upper (lower) number in curly braces representing the `no-flip' (`flip') probability for the emitted photon being in the same (different) polarisation state as the incoming photon. We thus have the cross section ratio
\bea
  \frac{\sigma_{\trm{D}}}{\sigma^{2\to 2}} \sim \frac{Z^2}{\eta} \; . \label{eqn:sigDto22}
\eea
Comparing the cross sections (\ref{eqn:sigDel}) and (\ref{eqn:sig.11}), or the ratios (\ref{eqn:sigDto22}) and (\ref{eqn:sig1to1foc}), we see that the $Z^{4}\alpha^2$  factor in Delbr\"uck scattering is replaced with the factor $\xi^{2}\Phi/\alpha$ in a classical background wave of intensity $\eta$ and phase duration $\Phi$. The low-energy Delbr\"uck scattering cross-section (\ref{eqn:sigDel}) assumes $Z\alpha \ll 1$ \cite{Milstein:1994zz} while the low-energy plane wave cross section (\ref{eqn:sig.11}) assumes small strong-field parameter, $\chi \equiv \xi\eta\ll 1$ (low energy and intensity). Note that the enhancement of the cross-section in the plane wave case can be much larger than for Delbr\"uck scattering due to the larger scattering  volume (measured by phase duration, $\Phi$) compared to a Coulomb centre. (See, however, suggestions for how to observe photon scattering in the superposition of multiple Coulomb fields \cite{Ahmadiniaz:2020kpl} and examples of the effect of multiple Coulomb fields in oriented crystals \cite{Wistisen:2017pgr}.)
\newline

In the monochromatic limit, the cross-section can be written as
\bea
  \sigma^{1\to 1} = \frac{\xi^{2}\Phi}{\alpha}\frac{1}{2\pi}
  \left(\frac{2\alpha r_{e}}{45}\right)^2 \left(\frac{\omega}{m}\right)^4\begin{Bmatrix} 11^{2} \\3^{2}\end{Bmatrix}\,, \label{eqn:sig1to1PW}
\eea
where $\Phi=\omega_{\vkap}T$ and $T$ is a divergent time factor\footnote{The different numerical prefactors in (\ref{eqn:sig.11}) and (\ref{eqn:sig1to1PW}) can be attributed to different definitions of $A$ and $T$ in terms of the focussing parameters. These differences arise if the plane-wave limit is taken at the level of the potential (as we have chosen), or the squared potential (as occurs in the definition of the flux).}. When the strong-field parameter $\chi=\xi\eta$ is not small, the cross-section no longer grows linearly with $\xi^{2}\Phi/\alpha$ and one must invoke the methods of strong-field QED to calculate the cross-section (for reviews, see e.g. \cite{ritus85,DiPiazza:2011tq,Narozhny:2015vsb,Fedotov:2022ely}). All orders of interaction between the virtual pair in the fermion loop and the background must be included. This can be achieved by calculating the $1\to1$ process where the fermions of the pair are replaced by `dressed' Volkov fermions solving the Dirac equation in a plane-wave background. If $ \eta \lesssim 2$, the probability of a photon flipping helicity in a linearly-polarised plane wave can be well-approximated by the locally constant field approximation \cite{dinu14a,King:2023eeo}. The helicity-flip cross-section in a monochromatic wave is then given by \cite{Heinzl:2006pn}
\bea
  \sigma^{1\to 1}_{+-} &=& \left(\frac{2\alpha r_{e}}{15}\right)^2 \left(\frac{\omega}{m}\right)^4 \nn \\ 
  && \!\times\frac{\xi^{2}\Phi}{2\pi\alpha}\left[1+\frac{32}{21}\xi^{2}\left(\frac{\omega}{m}\right)^4+\frac{1600}{143}\xi^{4}\left(\frac{\omega}{m}\right)^8+\cdots\right]\!.\nn \\
\label{eqn:sig1to1PWc}
\eea
The terms in square brackets remain a correction, as long as $\chi = 2\xi\omega^2/m^{2} \ll 1$, but at larger values of $\chi$ resum to an integral representation \cite{ritus85}. 
\newline 

Photon-photon scattering employing real rather than virtual photons has recently seen a renaissance: new experiments have been performed at MeV CM energies \cite{Watt:2024brh}, following suggestions to improve previous experimental bounds on the cross section \cite{Sangal:2021qeg}. In addition, there is theoretical work on how to measure the process when the cross-section becomes non-perturbative in the coupling to the classical field, assuming $\chi = O(1)$ \cite{king16,Bragin:2017yau,Macleod:2023asi}. In view of these developments, it seems useful to directly compare the cross-section for the vacuum and coherently-enhanced photon scattering channels.

\begin{figure}[h!!]
    \centering
\includegraphics[width=8.4cm]{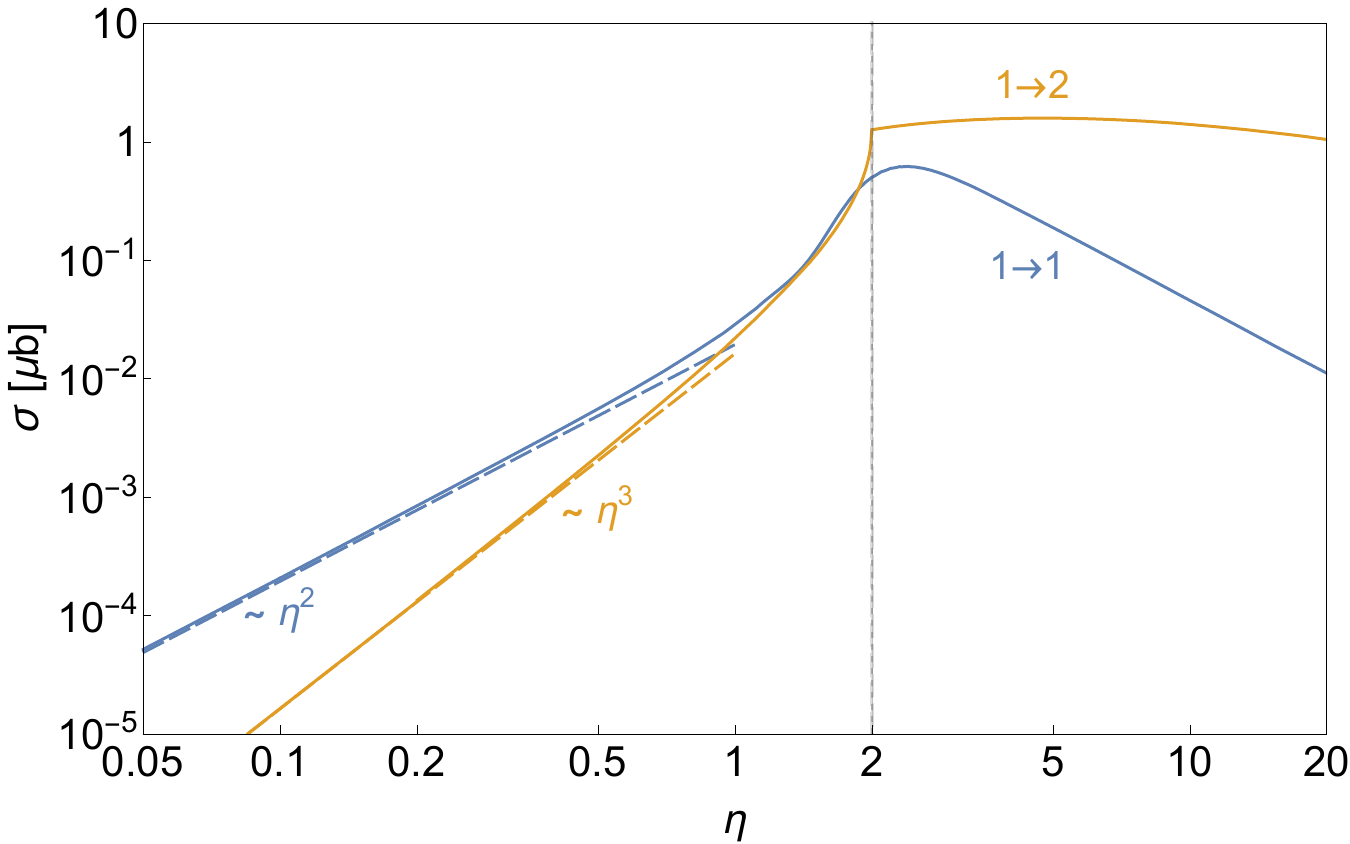}
\caption{Comparison of the total cross-section for the $1\to 2$ process, which is independent of the background field parameters, and the cross-section for helicity flipping in the $1 \to 1$ process in a linearly polarised plane wave with intensity parameter $\xi=0.1$. The grey vertical gridline corresponds to the threshold for the inelastic process (linear Breit-Wheeler pair creation). The dashed lines are the low-energy results in  (\ref{eqn:sig2to2a}) (for $1\to2$) and (\ref{eqn:sig1to1PW}) (for $1\to1$) with the pulse-length parameter $\Phi$ calculated for a sine-squared pulse as described in the text.} \label{fig:numericalResult}
\end{figure}

In the comparison, we take the $1\to 2$ process to be identical to the $2\to 2$ process, employing the relation (\ref{eqn:M.12.M.22}). We calculate the answer numerically using the integrals given in \cite{detollis64,detollis65,detollis71}.
For the $1\to 1$ process, we choose the classical field as a plane wave of the form (\ref{eqn:FmunuSimp}) such that the gauge potential has a sine-squared envelope and compact support,
\[
  A^{\mu}(\vphi) = m \xi \eps^{\mu} \sin^{2}\left(\frac{\vphi}{2N}\right)\,\cos\vphi; \qquad 0 \leq \vphi \leq 2N\pi \; ,
\]
where $\varphi = \kappa \cdot x$. The pulse phase duration, $\Phi$ is calculated using the mean-squared profile,
\bea
  \Phi = \int_{0}^{2N\pi} \sin^{4}\left(\frac{\vphi}{2N}\right)\,\cos^{2}\vphi~ d\vphi = \frac{3 N \pi}{8}.
\eea
We employ the numerical methods of \cite{King:2023eeo}, choosing $\xi=0.1$ and a linearly polarised background. Results for the cross sections are plotted in \figref{fig:numericalResult}. One can clearly see the differences in scaling at low energies, $\eta \ll 1$, where  $\sigma^{1\to1} \sim \eta^2 \sim (\omega/m)^{4} $ and $\sigma^{2\to2} \sim \eta^3 \sim (\omega/m)^{6} $, recall (\ref{eqn:sig.scaling}). At sufficiently high energies the $1\to 2$ process is dominant: when $\eta \gg 1$ (but $\xi \ll 1$) the numerical results suggest that $\sigma^{1\to2} \sim \eta^{-1} \sim (\omega/m)^{-2}$, whereas $\sigma^{1\to1}_{+-} \sim \eta^{-2} \sim (\omega/m)^{-4}$. If the intensity of the classical field is increased, the $1 \to 1$ cross-section will become larger, but also the shape around the pair threshold region, $2 < \eta < 2(1+\xi^2)$, can change \cite{King:2023eeo}. Eventually at a high-enough energy, the $1\to 2$ process dominates.
\newline 

\subsubsection{Vacuum birefringence}

Given the probability $\tsf{P}^{1\to1}$ we can link the intensity form factor to vacuum birefringence. Let us suppose that a photon in helicity state $\ket{\eps_{\pm}} = [\ket{\eps_{1}}\pm i\ket{\eps_{2}}]/\sqrt{2}$ probes a plane wave classical field that is polarised in e.g. the $\eps_{1}$ direction. Since the vacuum refractive indices in $\eps_{1}$ and $\eps_{2}$ directions differ (i.e. the vacuum is birefringent), a phase difference, $\Delta$, develops between the helicity states. In consequence, the probability for helicity flipping, $\tsf{P}^{1\to1}_{+-}$, can be written as the overlap 
\bea
  \tsf{P}_{+-} &=& \big|\proj{\eps_{+}(0)}{\eps_{-}(\vphi)}\big|^2 = \frac{1-\cos\Delta}{2} \; , \nn \\
  \Delta &=& -\frac{(\ell^{0})^2}{\vkap\cdot \ell}\int d\vphi  \, [n_{2}(\vphi)-n_{1}(\vphi)] \; ,\label{eqn:PflipRefInd}
\eea
where $n_{j}$ is the vacuum refractive index for photons polarised parallel to the $j$ direction and $\varphi = \vkap \cdot x = \omega_\vkap x^-$. For $\Delta \ll 1$, we see $\tsf{P}_{+-} \approx (\Delta/2)^2$. The phase integral in the intensity form factor can be isolated by introducing a partial Fourier transform $\bar{\chi}$ via
\bea 
  \chi(q) &=&  \int\frac{d\vphi}{2\omega_{\vkap}} \bar{\chi}(\vphi;q) \nn \; , \\
  \bar{\chi}(\vphi;q) &=& \int d^{2}x^{\perp}\,dx^{+}\,F^{2}(x) \mbox{e}^{iq\cdot x} \; , 
\eea
where we have once again assumed that the tensor structure of the field is independent of co-ordinates, i.e., $F^{\mu\nu} = F(x) \eps^{\mu\nu}$ as in the previous section. In this case, the flip probability becomes
\bea 
  \tsf{P}^{1\to1}_{+-} = V\int \frac{d^{3}\pmb{\ell'}}{(2\pi)^3} \Bigg|c^{\mu\nu\rho\sigma}\mathfrak{M}_{\mu\nu\rho\sigma} \int\frac{d\vphi}{2\omega_{\vkap}} \bar{\chi}(\vphi;q)\Bigg|^{2} \; . \nn \\
  \label{eqn:PflipRefDev}
\eea
To compare with the refractive index approach, we assume $F = F(\vphi)$ i.e., the background represents a plane wave, which allows the phase integral in (\ref{eqn:PflipRefDev}) to be isolated. Performing the integrations, we find
\bea 
  \tsf{P}^{1\to1}_{+-} = \Bigg|\frac{(2V)^{2}\ell^{0}}{2\,\omega_{\vkap}\,\vkap\cdot \ell}\, \left[\mathfrak{M}^{2\to2}_{+-}\right]_{1\to 1}\, \int d\vphi \,F^{2}(\vphi) \Bigg|^{2}
  \label{eqn:PflipRefDev2}
\eea
where the prescription on the $2\to2$ scattering amplitude is the same as in (\ref{eqn:Qlp1}, but now for a photon helicity flip. Using \eqnref{eqn:M1to1squared} and equating (\ref{eqn:PflipRefInd}) and (\ref{eqn:PflipRefDev2}), we recover the known result \cite{king15a},
\bea 
  |n_{2}(\vphi)-n_{1}(\vphi)| = \frac{2|\bar{c}_{2}-\bar{c}_{1}|\alpha}{45\pi} \left(\frac{\ell^{-}}{\ell^{0}}\frac{F(\vphi)}{\Ecr} \right)^2,
\eea
employing the critical (`Schwinger') field strength  $\Ecr=m^2/e$ and the dimensionless low-energy constants, $\bar{c}_{j} = 45m^4 c_{j}/2\alpha^2$, i.e., $\bar{c}_{1}=4$ and $\bar{c}_{2}=7$.

\subsubsection{Circularly-polarised monochromatic background}

A final application of the $1\to1$ probability is provided by a monochromatic classical field with circular rather than linear polarisation. This example is interesting for two reasons: (i) It cannot be calculated using the Heisenberg-Euler Lagrangian (which was derived for constant fields) since the instantaneous polarisation of a circularly-polarised field is \emph{not} constant in space and time. (ii) The low-energy asymptotics for circular polarisation differ from those of a linearly-polarised (hence constant polarisation) classical wave. It was recently shown \cite{King:2023eeo} that the low-energy limit of the probability $\tsf{P}^{1\to1}_{1\,2}$ for flipping the linear polarisation of the photon e.g.\ from state 1 to state 2 in a \emph{circularly polarised} background is
\bea 
  \tsf{P}^{1\to1}_{1\,2} = \left(\frac{2\alpha}{315\pi} \eta^2\int \xi^2(\vphi)d\vphi \right)^{2}.
\eea
For a monochromatic background, this corresponds to the cross-section
\bea 
  \sigma^{1\to1}_{1\,2} = \frac{\xi^{2}\Phi}{2\pi\alpha}\,\left(\frac{8\alpha r_{e}^{2}}{315}\right)^2\,\eta^{4},
\eea
such that $\sigma^{1\to1}_{1\,2} \sim \eta^4 \sim (\omega/m)^8$. This should be compared to the cross-section for photon helicity-flipping in a \emph{linearly-polarised} background, recall (\ref{eqn:sig1to1PWc}), which scales like $\sim (\omega/m)^4$. Note that the dependency on the field strength ($\sim \xi^2$) is the same for both polarisations: it cannot be described by higher-order terms in the Heisenberg-Euler Lagrangian without derivative corrections.

\section{Cross-section of 0-to-1 photon scattering}

The final light-by-light scattering process to be discussed is  the $0\to 1$ process of photon emission from a classical background (sometimes referred to as `vacuum emission' \cite{Karbstein:2019oej}). This can be defined in analogy with the $1\to 0$ process of (\ref{eqn:T-gamma-laser1t0}),
\bea \label{eqn:T-gamma-laser0t1}
  T_{fi}^{0 \to 1} &\sim&   \braket{\gamma}{f^{\mu\nu}(x)}{0} \, c_1 \chi^{(3)}_{\mu\nu}(q)  \nn \\
  &+& \braket{\gamma}{\fdual^{\mu\nu}(x)}{0}\left[c_{2} \, ^\ast\!\chi^{(3)}_{\mu\nu}(q)+c_{3}\chi^{(3)}_{\mu\nu}(q)\right],
\eea
where $q=\ell'$. We have defined the tensorial form factor 
\bea
  \chi_{\mu\nu}^{(3)} (q) &=& \int d^4 x \, \mbox{e}^{i q \cdot x} \, F^{\rho\sigma}(x)F_{\mu\nu}(x) F_{\rho\sigma}(x)\nn \\
  &\equiv& \tilde{F}^{\rho\sigma} \ast \tilde{F}_{\mu\nu} \ast \tilde{F}_{\rho\sigma} (q)  \;  \label{eqn:FF0t1a}
\eea
and $^{\ast}\!\chi^{(3)}$ is the same with one contracted $F$ tensor replaced by its dual. To calculate the probability, we integrate over the 3-momentum of the single emitted photon,
\bea
  \tsf{P}^{0\to 1} = V \int \frac{d^{3}\pmb{\ell'}}{(2\pi)^{3}}\,|T_{fi}^{0\to1}|^{2},
\eea
where $T_{fi}^{0\to1} = \mathfrak{M}^{0\to1}_{\mu\nu} \,  \chi^{(3)\,\mu\nu}$. Considering again backgrounds with constant tensor structure, $\epsilon_{\mu\nu}$, we can write
\bea
  \chi^{(3)}_{\mu\nu}(q) = \epsilon_{\rho\sigma} \epsilon^{\rho\sigma} \epsilon_{\mu\nu} \chi^{(3)}(q),
\eea
and analogously for $^{\ast}\!\chi^{(3)}_{\mu\nu}(q)$. This turns the scattering amplitude into
\bea
  T_{fi}^{0\to1} = \frac{(2V)^{3}}{\omega_{1}\omega_{2}\omega_{3}}\chi^{(3)}(q)\left[\mathfrak{M}^{2\to2}\right]_{0\to1}
\eea
where $\omega_{1,2,3}$ are the frequencies of the wave fields in $\chi^{(3)}(q)$. We note that the nonlinear form factor (\ref{eqn:FF0t1a}) has a contraction $F\cdot F$ and therefore, for a single focussed laser pulse, is approximately zero. in order to have something measurable in experiment, we consider a scenario of multiple, colliding, laser pulses. There has been recent interest \cite{Gies:2017ezf,King:2018wtn,Aboushelbaya:2019ncg,Berezin:2024fxt} in looking at a collision of three laser pulses, because of the kinematic control it gives over the final photon. Suppose then that the vector potential for the classical field is chosen to be the sum of three monochromatic waves (chosen for clarity of exposition):
\bea 
  A_{\mu}(x) &=& \sum_{j=1}^{3}A^{(j)}_{\mu}(x) \nn \\
  A^{(j)}_{\mu}(x) &=& \frac{m\xi_{j}}{e} \epsilon^{(j)}_{\mu}\,\cos \left(\vkap_{j} \cdot x\right).
\eea
There are various channels contributing to $\hat{\chi}_{\mu\nu}$; some involve Fourier modes from just two of the waves, but we concentrate here on the three-beam interaction, for which
\bea 
\chi^{(3)}(q) = \frac{F_{1,0}F_{2,0}F_{3,0}}{8}\!\!\sum_{\lambda_{1},\lambda_{2},\lambda_{3}}\delta\left(q+\lambda_{j}\vkap_{j}\right) \; , \nn \\
\eea
where $F_{i,0}$ is the field strength of beam $i$ and in the argument of the delta function, a sum over $j=1,2,3$ is implied, and $\lambda_{j} \in\{-1,1\}$. Assuming (for simplicity) the typical case that channels do not overlap in momentum space, the contribution to the probability is
\bea 
  \hspace{0.5cm}\tsf{P} &=& \frac{V^{3}\left(F_{1,0}F_{2,0}F_{3,0}\right)^{2}}{2^{3}\omega_{1}\omega_{2}\omega_{3}}\int \frac{d^{4}q}{(2\pi)^3} 2q^{0}\left|\left[\mathfrak{M}^{2\to2}\right]_{0\to1}\right|^2 \nn \\
  && \qquad\qquad\times \theta\left(q^{0}\right)\delta\left(q^2\right)\sum_{\lambda_{1},\lambda_{2},\lambda_{3}} \delta^{2}(q+\lambda_{j}\vkap_{j}) \label{eqn:P.0.1} \nn\\
\eea
The scattered photon is real with its momentum on mass-shell (and hence can propagate to the detector) whenever the beam 4-momenta obey
\bea
  \lambda_{1}\lambda_{2}\,\vkap_{1}\cdot\vkap_{2} + \lambda_{1}\lambda_{3}\,\vkap_{1}\cdot\vkap_{3} + \lambda_{2}\lambda_{3}\,\vkap_{2}\cdot\vkap_{3} = 0. \nn \\
\eea
In the following, we distinguish two important cases that are kinematically allowed.
\bigskip

\subsection{Forward scattering}

Forward scattering occurs if $\vkap_{2} = \vkap_{3} = \vkap'$ and corresponds to the case where the classical background comprises only two colliding waves, such that $\lambda_{2} + \lambda_{3} = 0$, i.e., one Fourier mode must be absorbed from, and one Fourier mode emitted back into the wave. This is analogous to the $1\to1$ process of the previous section, where the incoming photon is replaced by a Fourier mode of the classical field, and an integral over the wave packet of incoming Fourier modes must be applied. Therefore, for the $0\to 1$ process in a monochromatic wave, this channel gives a cross-section analogous to the one for the $1\to1$ process in a monochromatic background, but with the difference that the incoming flux of \emph{two} classical waves is already included. Using the definition (\ref{eqn:fluxA}), the incoming flux of the two waves is
\bea
  J &=& J_{11}+J_{22}+2J_{12} \nn \\
  &=&  \frac{F_{1,0}^2}{2\omega_{1}}A_{1}T_{1}+\frac{F_{2,0}^2}{2\omega_{2}}A_{2}T_{2} \nn \\ 
  && - \delta_{\omega_{1},\omega_{2}} \epsilon_{1}\cdot\epsilon_{2}\cos(\psi_{12})\frac{F_{0,1}F_{0,2}}{\omega_{1}}A_{12}T_{12},
\eea
where $\psi_{12}$ is the collision angle in the lab frame ($\psi_{12}=\pi$ for a head-on collision), $\epsilon_{j}$ are the polarisation vectors of the waves in the lab. The volumetric factors, areas $A_i$ and time durations $T_i$, have been labelled according to their origin ($i = 1,2,12$ for beam 1, beam 2 and their overlap, labelled 12). (The overlap is only non-zero if the frequencies overlap, which for monochromatic waves means the frequencies must be identical.) The flux leads to a cross section,
\bea 
  \sigma^{0\to1} = \frac{2\xi_{1}\xi_{2}^{3}}{(4\pi \alpha)^3}
  \left(\frac{2\alpha r_{e}}{45}\right)^2 \left(\frac{\omega}{m}\right)^4 G_{12}\begin{Bmatrix} 11^{2} \\3^{2}\end{Bmatrix}\,, \nn \label{eqn:sig0to1PWa}\\
\eea
where we have assumed that intensities $\xi_{1}$ and $\xi_{2}$ characterise probe and background, respectively (i.e. $\xi_{2}\gg \xi_{1}$).  We have also defined a geometric factor, $G_{12}$, that depends on the relative intensities, waists and durations of the two colliding beams:
\bea 
  G_{12} = && \omega_{1}\omega_{2}\xi_{1}\xi_{2}\mathcal{A}_{12} \mathcal{T}_{12}^{2} / 2\left[ \omega_{1}\xi_{1}^2 A_{1}T_{1} + \omega_{2}\xi_{2}^{2}A_{2}T_{2} \right. \nn \\[5pt]
  && \left.  -2 \delta_{\omega_{1},\omega_{2}} \epsilon_{1}\cdot\epsilon_{2} \cos(\psi_{12})\omega_{1}\xi_{1}\xi_{2}A_{12}T_{12}\right] 
  \; , 
  \label{eqn:G12a}
\eea
with $\epsilon_{j}$ the polarisation vectors in the CM frame. 
We note that the volumetric factors $\mathcal{A}$ and $\mathcal{T}$ are, in general, \emph{different} from the factors to $A_i$ and $T_i$: The former originate from the form factor, which contains an integral over three waves multiplied together, whereas $A_i$ and $T_i$ stem from the flux, which is an integral over two waves multiplied together. 

As an example, let us consider the head-on collision of two paraxial Gaussian beams using (\ref{eqn:F-paraxial}), and assume them to be perpendicularly-polarised with respect to each other. In this case,  \eqnref{eqn:fluxGauss1} implies the flux
\bea
  J = \frac{m^{2}}{4\pi \alpha} \left(\frac{\pi}{2}\right)^{3/2}\xi^{2}\omega\tau w_{0}^{2},
\eea
from which we infer that 
\bea
  A_{j}T_{j} \sim w_{0,j}^{2} \tau_{j} \; , \quad 
  j = 1,2 \; .
  \label{eqn:AT.FLUX}
\eea
To analyse the volumetric factor $\mathcal{A}_{12}\mathcal{T}_{12}^{2}$ appearing in the numerator of (\ref{eqn:G12a}), we calculate the form factor integral (\ref{eqn:FF0t1a}) and find
\bea
  \chi^{(3)}(q) = \frac{\pi^{2}}{2^{3/2}}\tau_{1}\tau_{2}w_{122}^{2}F_{1,0}F^{2}_{2,0} \exp f_{2}(q) \; ,
  \label{eqn:FF.01}
\eea
where we have introduced the exponent
\bea 
  f_{2}(q) &=& -\frac{1}{4} \left[ \tau_{1}^{2}\left(\omega_{1}+\frac{q^{0}+q^{3}}{2}\right)^{2}  \! + w_{122}^{2} \mbf{q}_{\perp}^{2}\right], 
\eea
and an effective waist area,
\bea
  w_{122}^{2} &=& \frac{w_{0,1}^{2}w_{0,2}^{2}}{2w_{0,1}^{2}+w_{0,2}^{2}}.
\eea
Comparing with (\ref{eqn:G12a}) we infer that
\bea
  \mathcal{A}_{12} \mathcal{T}_{12}^2 \sim w_{122}^2 \tau_1 \tau_2 
  \; ,
\eea
which clearly differs from the $A_{j}T_{j}$ factors in (\ref{eqn:AT.FLUX}) stemming from the flux. 

\subsection{Frequency shifting}

The phenomenon of frequency shifting can occur if the two monochromatic beams and the incoming photon form an orthogonal dreibein when they collide. In this case, the delta function in (\ref{eqn:P.0.1}) has non-empty support if
\bea \label{eqn:FS.COND}
  \frac{1}{\lambda_{1}\omega_{1}} + \frac{1}{\lambda_{2}\omega_{2}} +  \frac{1}{\lambda_{3}\omega_{3}} = 0.
\eea
For the particular choice that Fourier modes $\vkap_{2}$ and $\vkap_{3}$ are both absorbed (so $\lambda_{2}=\lambda_{3}=-1$), (\ref{eqn:FS.COND}) can only be fulfilled if a Fourier mode with momentum $\vkap_{1}$ is emitted, i.e., $\lambda_{1} = 1$. Defining the frequency ratios $\nu_{j} = \omega_{j}/\omega_{1}$, realisation of this channel requires
\bea
\frac{1}{\nu_{2}} +  \frac{1}{\nu_{3}} = 1.
\eea
A common choice explored previously \cite{Lundin2006PRA,lundstroem_PRL_06,Gies:2017ezf,King:2018wtn,Aboushelbaya:2019ncg,Berezin:2024fxt} is to frequency-double two lasers, hence to choose
\bea 
  \nu_{2} = \nu_{3} = 2. \label{eqn:nuChoice1}
\eea
In this case, the kinematics for the $0\to 1$ channel is
\bea
  \vkap_{2}+\vkap_{3}=\ell' + \vkap_{1},
\eea
and the Mandelstam variables become 
\bea
  s &=& (\vkap_{2}+\vkap_{3})^2 = 4\omega^2 = 2\omega_{2}\omega_{3} \; , \\ 
  t &=& (\vkap_{2}-\vkap_{1})^2 = 2\omega_{1}\omega_{2} \; , \\
  u &=& (\vkap_{1}-\vkap_{3})^{2} = 2\omega_{1}\omega_{3} \;. 
\eea
It then follows from \eqnref{eqn:nuChoice1} that $t=s/2$ and $u=t/2$ corresponding to a CM emission angle of $\theta=\pi/2$. This is quite different from the $1\to1$ case in a plane wave, for which $t=s$ and $u=0$, corresponding to $\theta=0$. As a result, we find slightly different numbers for the cross-section, defined as
\bea
  \sigma^{0\to1} = \frac{1}{\flux}\frac{V}{T}\tsf{P}^{0\to1},
\eea
where $\flux$ is the incoming flux from the classical field, 
\bea
  \flux = \sum_{j=2}^{3}\flux_{j} =\sum_{j=2}^{3}\frac{F_{j,0}^{2}}{2\omega_{j}}\,A_{j}T_{j},
\eea
As before,  $A_{j}$ and $T_{j}$ are divergent area and time factors that depend on the beam details. We note there is no flux from the interference term between the beams as their directions are mutually orthogonal. Using the same designation for CM quantities as in \eqnref{eqn:poldef}, we find the polarisation-resolved cross sections,
\bea
  \begin{Bmatrix} \sigma[(2_{\parallel},2_{\parallel})\to(1_{\parallel},3_{\parallel})] \\ \sigma[(2_{\parallel},2_{\parallel})\to(1_{\perp},3_{\parallel})] \\
  \sigma[(2_{\parallel},2_{\perp})\to(1_{\parallel},3_{\perp})] \\
  \sigma[(2_{\parallel},2_{\parallel})\to(1_{\perp},3_{\perp})] 
  \end{Bmatrix} = \sigma^{(0)}\begin{Bmatrix} (6\bar{c}_{1})^{2} \\ 0 \\ (\bar{c}_{1}-5\bar{c}_{2})^{2}\\
  2^2(2\bar{c}_{1}-\bar{c}_{2})^{2} \; ,
\end{Bmatrix}, \label{eqn:sig0to1PWa}\nn 
\eea
where the integer in round brackets label the harmonic number of the frequencies involved. The universal prefactor on the right is
\bea
  \sigma^{(0)}=\frac{\xi_{1}^{2}\xi_{2}\xi_{3}}{4\pi \alpha^{2}}\frac{1}{2\pi}
  \left(\frac{\alpha r_{e}}{180}\right)^2 \left(\frac{\omega}{m}\right)^4\,G_{23} \; .
\eea
Once again, we have defined a geometric factor, $G_{23}$, which depends on the relative waists and durations of the three beams,
\bea 
  G_{23} = \frac{1}{4}\frac{\xi_{2}\xi_{3}\mathcal{A}_{123}\Phi_{123}}{\xi_{2}^{2}A_{2}\Phi_{2}+\xi_{3}^{2}A_{3}\Phi_{3}},
\eea
with $\Phi_{123} = \ell'^0 T_{123}$ and $\mathcal{A}_{123}$ denoting the phase and area factors arising from the beam overlap integral in (\ref{eqn:P.0.1}). For the given set-up, $\ell'^{0} = 3\vkap^{0}$, where $\vkap^{0}$ is the fundamental frequency, while the CM frequency occurring in the cross-section is $\omega = \sqrt{2}\vkap^{0}$. For QED values, the cross-sections (\ref{eqn:sig0to1PWa}) become
\bea
  \begin{Bmatrix}
  \sigma[(2_{\parallel},2_{\parallel}) \to (1_{\parallel},3_{\parallel})] \\
  \sigma[(2_{\parallel},2_{\parallel}) \to(1_{\perp},3_{\parallel})] \\
  \sigma[(2_{\parallel},2_{\perp}) \to (1_{\parallel},3_{\perp})] \\
  \sigma[(2_{\parallel},2_{\parallel}) \to (1_{\perp},3_{\perp})] 
  \end{Bmatrix} 
  = \sigma^{(0)}\begin{Bmatrix} 24^{2} \\ 0  \\ 31^{2} \\ 2^{2}
  \end{Bmatrix}.
\eea

\section{Two-photon pair annihilation}

Pair annihilation in a classical background has been studied over the years in a number of papers \cite{nikishov64,ritus85,Ilderton:2011ja,Tang:2019ffe,Bragin:2020akq}. Here, we add to these studies by applying the idea of a coherent enhancement form factor to pair-annihilation into two photons. This is a \emph{linear} process as it can happen in standard (low-intensity) QED. We will first review the calculation of precisely this QED process which, in our present language, is the $2 \to 2$ vacuum process. In a second step, we will adhere to the philosophy of this paper and replace one of the outgoing photons with a classical background field line. This amounts to calculating the $2\to1$ process.

\subsection{Cross-section of 2-to-2 pair-annihilation}

To acquire the diagram for pair annihilation into two photons (and its conjugate), the  loop diagram for photon-photon scattering can be cut and a crossing transformation employed. 
As a result, the $2 \to 2$ process, $e^{+}(q) e^{-}(p) \to \gamma_1 (\ell_1) \gamma_2 (\ell_2)$, has the second-order scattering amplitude,
\bea
  \tmat_{fi}^{2 \to 2} = -e^2\!\!\int \!\!d^4x \, d^4 y \,  \braket{\gamma_1 \gamma_2}{j_\mu(x) A^\mu(x) j_\nu(y) A^\nu(y)}{e^+ e^-} \!.\nn \\
\eea
Disentangling the transition matrix elements and using translation invariance exhibits 4-momentum conservation and two invariant amplitudes, depicted in \figref{fig:annihilation1} (left panel),
\bea
  \tmat_{fi}^{2 \to 2} &=& (2\pi)^4 \delta(p + q - \ell_1 - \ell_2 )i\mathcal{M}^{2\to2} \; , \\
  \mathcal{M}^{2\to2}&=&ie^2~ \braket{\gamma_1}{\bar{\psi} \slashed{A}}{e^+} \, \frac{1}{\slashed{\ell}_1 - \slashed{q}-m} \, \braket{\gamma_2}{\slashed{A} \psi}{e^-} \nonumber \\
  && + (1 \leftrightarrow 2) \; , \label{eqn:M.22.PA1}
\eea
with all field operators evaluated at the origin, $x=y=0$. Introducing the standard one-particle wave functions,
\bea
  \psi_{p}(x) &=& \braket{0}{\psi(x)}{e^-(p)} = \frac{u_{p}}{\sqrt{2Vp^{0}}} \, \mbox{e}^{-ip\cdot x} \; , \nonumber \\  \bar{\psi}_{q}(x) &=& \braket{0}{\bar{\psi}(x)}{e^+(q)} = \frac{\bar{v}_{q}}{\sqrt{2Vq^{0}}} \, \mbox{e}^{-iq\cdot x} \; , \nonumber \\
  A^{\ast}_\mu(x) &=& \braket{\gamma}{A^\mu}{0}= \frac{1}{\sqrt{2V\ell^{0}}} \varepsilon^*_\mu  \, \mbox{e}^{i\ell \cdot x} \; , 
\eea
and using the Fourier representation of the fermion propagator, the amplitude (\ref{eqn:M.22.PA1}) becomes
\bea 
 \mathcal{M}^{2\to2}&=& \frac{1}{\sqrt{(2V)^{4}p^0q^0\ell^0 \ell^{\prime\,0}}}\left[\frac{\bar{v}_{q}\slashed{\eps}^{\prime\,\ast}\left(\slashed{p}-\slashed{\ell}+m\right)\slashed{\eps}^{\ast}u_{p}}{p\cdot l} \right. \nn \\ && \qquad \left.+\frac{\bar{v}_{q}\slashed{\eps}^{\ast}\left(\slashed{\ell}-\slashed{q}+m\right)\slashed{\eps}^{\prime\,\ast}u_{p}}{q\cdot l}\right],
\eea
where $\eps$ and $\eps'$ are the polarisation vectors of the two emitted photons with momenta $\ell$ and $\ell'$, respectively. Employing four-momentum conservation, the probability takes on the compact form
\bea 
  \tsf{P}^{2\to2} = V^{3}T\int \frac{d\Omega_{\ell}}{(2\pi)^{2}}~\omega_{\ell,\ast}^2|\mathcal{M}^{2\to2}|^2,
  \label{eqn:P2to2aba}
\eea
where $d\Omega_{\ell}$ is the unit solid angle in the outgoing photon momentum $\ell$. The energy $\omega_{\ell,\ast}$ is fixed by four-momentum conservation; in the CM frame it is equal to the CM energy $\omega_{\ell,\ast}=\omega$ where $\omega^2=(p+q)^2/4$.

\medskip

We define the annihilation cross-section as
\bea \label{eqn:sigPairAnn}
  \frac{d\sigma^{2\to2}}{d\Omega} = \frac{V}{T}\frac{1}{v_{\trm{rel.}}}\frac{d\tsf{P}^{2\to2}}{d\Omega}; \quad v_{\trm{rel.}} = \frac{p^0 q^0}{\sqrt{(p\cdot q)^2-m^4}}
  \; , \nn \\ 
\eea
and the integral is then calculated in the electron CM frame. Averaging over initial fermion spins and summing over photon polarisations, yields the cross-section \cite{greiner2013quantum}
\bea 
  \sigma^{2\to 2} &=& \frac{r_{e}^{2}}{8\pi}\left\{\frac{\sba+1}{\sba}\sqrt{\frac{1}{\sba(\sba-1)}} \right. \nn \\ && \left. +\frac{2\sba^2+2\sba-1}{2\sba^2}\ln\left(2\sba-1+2\sqrt{\sba(\sba-1)}\right)\right\}
  \; , \nn \\
\eea 
where we have introduced the normalised Mandelstam variable $\bar{s}=(p+q)^2/4m^2$, which is bounded from below by $\bar{s}>1$ which corresponds to the rest energy of the pair.

\subsection{Cross-section of 2-to-1 pair-annihilation}

\begin{figure}[h!!]
\centering
\includegraphics[width=8.4cm]{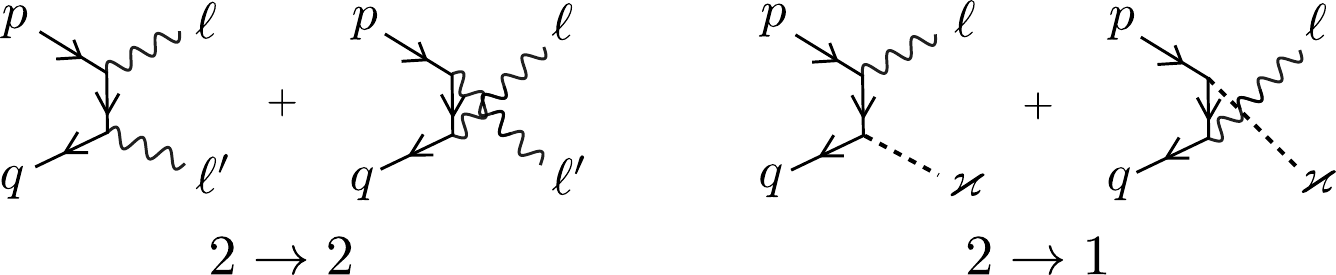}
\caption{Pair-annihilation diagrams for the $2\to2$ and $2\to1$ processes.} \label{fig:annihilation1}
\end{figure}

We now replace an outgoing photon by an external field, $A$, and hence consider the $2\to1$ pair annihilation process. We are thus left with a single outgoing photon, $\gamma(\ell)$ (see Figure.~\ref{fig:annihilation1}, right panel). As we will see in a moment, the resulting kinematics differs substantially from that of the $2\to2$ process. Performing the replacement in the matrix elements, we find the scattering amplitude,
\bea
  \tmat_{fi}^{2 \to 1} &=& i\chi^{\mu}(Q) \mathcal{M}_{\mu}^{2\to1} \nn \\
  \mathcal{M}_{\mu}^{2\to1} &=&   -ie^2\braket{\gamma}{\bar{\psi} \slashed{A}}{e^+} \, \frac{1}{\slashed{\ell} - \slashed{q}-m}  \gamma_{\mu}\braket{0}{\psi}{e^-}
  \label{eqn:T.21.PA} \\
  && -ie^2 \braket{0}{\bar{\psi}}{e^+}  \, \gamma_{\mu}\frac{1}{\slashed{p} - \slashed{\ell}-m} \, \braket{\gamma}{\slashed{A} \psi}{e^-}. \nn \label{eqn:M.21.PA}\\
\eea
Here we have introduced a form factor, $\chi^\mu$, which is nothing but the Fourier transform of the gauge potential,
\bea 
  \chi_{\mu}(Q) = \int d^{4}x\, A_{\mu}(x)\,\mbox{e}^{iQ\cdot x} \equiv z_\mu^* (Q) 
\eea
and depends on the momentum transfer $Q = p+q-\ell$. By definition, this form factor agrees with the coherent state eigenvalue, $z_\mu^*$, and hence corresponds to a classical emission amplitude signalling (once again) coherent enhancement of the process. 
Evaluating the matrix elements in (\ref{eqn:M.21.PA}), the invariant amplitude becomes 
\bea 
 \mathcal{M}_{\mu}^{2\to1}&=& \frac{1}{\sqrt{(2V)^{3}p^0q^0\ell^0}}\left[\frac{\bar{v}_{q}\slashed{\eps}^{\ast}\left(\slashed{p}-\slashed{\ell}+m\right)\gamma^{\mu}u_{p}}{p\cdot l} \right. \nn \\ && \qquad \left.+\frac{\bar{v}_{q}\gamma^{\mu}\left(\slashed{\ell}-\slashed{q}+m\right)\slashed{\eps}^{\ast}u_{p}}{q\cdot l}\right].
\eea
For a plane wave classical field of standard form,
\bea
  eA^{\mu}(x) = \epsilon^{\mu} m\xi \, f(\vkap \cdot x),
\eea
with $\vkap\cdot x = \omega_{\vkap} x^{-}$, $\epsilon^{0}=0$, $\epsilon\cdot \vkap = 0$ and $\epsilon \cdot \epsilon=-1$, the form factor in (\ref{eqn:T.21.PA}) can be written as $\chi^{\mu}(Q) = \epsilon^{\mu}\chi(Q)$, where we have introduced the scalar form factor
\bea 
  \chi(Q) = 2\pi^{3}\frac{m\xi}{e \omega_{\vkap}}\,\delta_{\perp,+}(Q)\,\tilde{f}\left(\frac{Q_{-}}{\omega_{\vkap}}\right).   \label{eqn:chiQ}
\eea

Hence, for a plane wave background, the form factor sets three components of the momentum transfer, $q_+, q_1, q_2$, equal to zero. This fixes the same three components of the emitted photon and allows all the outgoing phase-space integrals to be performed with the result
\bea 
  \tsf{P}^{2\to1} &=& V\int \frac{d^{3}\pmb{\ell}}{(2\pi)^{3}}\big|\tmat^{2\to1}_{fi}\big|^2 \nn \\ &=& 
  2V^{3}\omega_{\vkap}\frac{\ell^{0}p^{0}}{\vkap \cdot \ell\, \vkap \cdot p}\bigg|\frac{m\xi}{e} \tilde{f}\left(\frac{2\,\bar{s}}{\eta_{\ell}}\right)\left[\mathfrak{M}^{2\to2}\right]_{2\to1}\bigg|^{2} \label{eqn:P2to1ann1}\nn \\
\eea
where $\eta_{\ell}=\eta_{p}+\eta_{q}=(p+q)\cdot \vkap/m^2$. The prescription on the matrix element corresponds to evaluating the $2\to2$ amplitude employing $2\to 1$ momenta and polarisations of the classical field, which amounts to the replacements
\bea
  \ell' \to \vkap; \quad \eps' \to \epsilon.
\eea
Comparing the probabilities for the $2\to2$ and $2\to1$ processes given by (\ref{eqn:P2to2aba}) and (\ref{eqn:P2to1ann1}), respectively, reveals that the kinematics are quite different: For the $2\to2$ process, the photons can be emitted in a range of angles. Instead, for the $2\to 1$ process, the kinematics, in particular the emission direction, is entirely fixed by the background.
\newline 

This may be further elucidated as follows. Suppose we define $r$ as the Fourier variable conjugate to external field phase, $\vphi=\omega_{\vkap}x^{-}$, i.e., 
\bea
  \tilde{f}(r) = \int d\vphi \, e^{ir\vphi} f(\vphi) \; , 
  \label{eqn:FOURIER.R}
\eea
for any function of phase, $f(\vphi)$. A monochromatic classical field with central frequency $\omega_{\vkap}$ then only has support when $r=1$. Momentum conservation in $2\to1$ pair annihilation requires the outgoing wave to have $r = r_{\ast} \equiv 2\bar{s}/\eta_\ell$. Calculating $\vkap \cdot p$ and $\vkap \cdot q$ and solving for the emission angle $\vartheta$ in CM co-ordinates implies that
\bea
  \cos \vartheta = \frac{\eta_{q}-\eta_{p}}{\eta_{\ell}\sqrt{1-1/\bar{s}}} .
\eea
The emission angle is thus completely determined by input parameters (i.e., momenta $p$, $q$ and $\ell$), unlike in the $2\to2$ process, where the emission angle is free to vary. This  is the same behaviour as found when comparing the $1\to1$ and $1\to2$ photon scattering processes in the previous section.
\newline 

The $2\to1$ cross-section, $\sigma^{2\to 1}$, can be defined by analogy with the $2\to2$ process, recalling (\ref{eqn:sigPairAnn}). For an \emph{unpolarised} emitted photon, one finds
\bea 
  \sigma^{2\to1} &=& 2\pi r_{e}^{2}\frac{\xi^{2}}{\alpha\eta_{\ell}\eta_{p}}\frac{p^0}{Tm^{2}}\big|\tilde{f}\left(r_{\ast}\right)\big|^2\frac{G}{\sqrt{\bar{s}(\bar{s}-1)}} \; , \nn \\
  G &=& \frac{\eta_{p}^{2}+\eta_{q}^{2}}{\eta_{p}\eta_{q}}+\frac{2}{r_{\ast}^{2}}\left[\frac{p\cdot \eps}{m\eta_{p}} - \frac{q\cdot \eps}{m\eta_{q}}\right]^{2}.
\eea
If we assume that the classical plane wave has a Gaussian envelope given by the profile function $f(\vphi) = \exp\left[-\vphi^2/\Phi^2\right]\cos \vphi$, the Fourier transformation (\ref{eqn:FOURIER.R}) becomes
\bea
  \tilde{f}(r) = \frac{\Phi \sqrt{\pi}}{2}\left\{\exp\left[-\frac{\Phi^2(r-1)^2}{4}\right]+(r\to -r)\right\}.\nn \\
\eea
This leads to the cross section
\bea 
  \sigma^{2\to1} &=& \frac{\pi^{2}r_{e}^{2}}{2}\frac{\xi^{2}\Phi}{\alpha \eta_{\ell}} \frac{G}{\sqrt{\bar{s}(\bar{s}-1)}}\nn \\
  && \times \left[\exp\left\{-\left[\frac{\Phi}{2}\left(\frac{2\bar{s}}{\eta_{\ell}}-1\right)\right]^{2}\right\}+(\bar{s}\to -\bar{s})\right]^2 \nn \\
  \label{eqn:SIGMA.21.PA}
\eea
where we used $p^{0}/p^{-}=T/L^{-}$ and $\omega_{\vkap}L^{-}=\Phi$. We note again how different the kinematic dependencies of the $2\to 1$ and $2\to 2$ annihilation cross-sections are: in the $2\to 1$ case, the classical field has supplied the momentum to enable the process in the first place. (The annihilation of a pair to a single photon is kinematically forbidden in vacuum.) Momentum conservation then fixes the corresponding momentum of the emitted photon. In the $2\to2$ process, photons can be emitted at a range of angles and energies. Comparing the scaling of the cross-sections, we see again that the field-assisted process introduces an overall coherent enhancement of $\sigma^{2\to 1}/\sigma^{2\to 2} \sim \xi^{2}\Phi/\alpha\eta_{\ell}$. If $\xi \not \ll 1$, higher orders of the interaction between the field and the pair must be taken into account, and the $\sigma^{2\to 1}$ enhancement takes on a more complicated form (see e.g. \cite{nikishov64,ritus85,Ilderton:2011ja,Tang:2019ffe,Bragin:2020akq}).
\newline 

The coherent enhancement of the annihilation cross section in a classical background field is not guaranteed automatically: the field momentum distribution centred around $\vkap$ must match the momenta of the annihilating electron and positron to avoid kinematic suppression. We can see this from the plane wave cross section (\ref{eqn:SIGMA.21.PA}), which requires $r_{\ast} = 2\bar{s}/\eta \approx 1$ in the exponents to escape exponential suppression. In an optical beam these requirements are difficult to fulfil: suppose we take an optical wavelength of  $800\,\trm{nm}$, equivalent to $\omega_{\vkap} = 1.55\, \trm{eV}$ as typical, then $\eta_{p}=1$ for a head-on collision with an electron of energy $80\,\trm{GeV}$, which exceeds currently available energies in laser labs by an order of magnitude. Since there should be some opening angle between the electron and positron, the energy requirement is likely to be even higher. However, if one instead considers an XFEL beam, thus a typical energy of $\omega_{\vkap}=10\,\trm{keV}$ \cite{Abela:77248}, then $\eta_{p}=1$ is reached in a head-on collision with an electron of energy $10\,\trm{MeV}$. The price to pay for this advantage, though, is a reduction of the intensity parameter $\xi\propto 1/\omega_{\vkap}$ and hence of the coherent enhancement.
\newline 

Let us discuss a concrete example. Starting from the general expression for the rescaled Mandelstam variable,
\bea 
  \sba = \frac{\eta_{l}^{2}}{4\eta_{p}\eta_{q}} + \frac{1}{4}\left(\sqrt{\frac{\eta_{p}}{\eta_{q}}}\frac{\mbf{q^{\perp}}}{m} - \sqrt{\frac{\eta_{q}}{\eta_{p}}}\frac{\mbf{p^{\perp}}}{m}\right)^2, \label{eqn:S.BAR}
\eea
we suppose that an electron and positron of equal energy collide with the laser at an angle $\psi$, where $\psi=0$ corresponds to a head-on collision.  Then $\mbf{p}^{\perp} = -\mbf{q}^{\perp}$ with $|\mbf{p}^{\perp}| =  |\mbf{p}| \sin\psi$ implying $\eta_{p} = \eta_{q} = \eta$ and $\eta_{l}=2\eta$. As a result, (\ref{eqn:S.BAR}) simplifies considerably,
\bea
  \sba = 1+ \frac{|\mbf{p}|^2\sin^{2}\psi}{m^{2}} \; ,
\eea
and $r_*$ becomes
\bea
  r_{\ast} = \frac{2\sba}{\eta_{p}} = \frac{2m^2\sba}{\vkap^{0}(p^0 + |\mbf{p}| \cos\psi)} \; ,
\eea
with $(p^0)^2 = |\mbf{p}|^2+m^2$. We assume that the electron and positron are highly relativistic, so the Lorentz parameter $\gamma \approx |\mbf{p}|/m \gg 1$. To evade kinematic suppression, we require $r_{\ast} \approx 1$, which gives a condition on the $\gamma$ factor of the electron and positron,
\bea
  \gamma \approx \frac{\omega_{\vkap}}{2m(1-\cos\psi)}\left[1\pm \left(1-\frac{4m^2(1-\cos\psi)}{\omega_{\vkap}^2(1+\cos\psi)}\right)^{1/2}\right],\nn \\
  \label{eqn:gammaEq1}
\eea
This only has real solutions if
\bea
  \cos\psi > \frac{4m^{2}-\omega_{\vkap}^2}{4m^{2}+\omega_{\vkap}^2}. \label{eqn:cosPsi1}
\eea
As a numerical example we adopt photons produced by the SASE 1 mode of the EU.XFEL, which have energies of $12.4\,\trm{keV}$ in $100\,\trm{fs}$ pulses of around $10^{12}$ photons (see Tab.\ 5.2.2 in \cite{Altarelli:2006zza}). To achieve $r_{\ast}\approx 1$, using the arguments above, one finds $\psi \lesssim \omega_{\vkap}/m$, corresponding to $\psi\lesssim 1.4^{\circ}$ and $\gamma\approx m/\omega_{\vkap}\approx 40$, i.e. $20\,\trm{MeV}$ electrons and positrons for which $\bar{s} \approx 2$ and $\eta_{p} \approx 2$. Using the EU.XFEL parameters, we assume that the SASE 1 x-ray beam can be focussed to $5\,\mu\trm{m}$ (as planned for upcoming vacuum birefringence experiments \cite{Ahmadiniaz:2024xob}), which implies a small intensity parameter of $\xi \approx 6 \times 10^{-5}$. For a full-width-at-half-maximum pulse duration of $30\,\trm{fs}$, so $\Phi \approx 3.5 \times 10^5$, the coherence factor turns out to be $\xi^{2} \Phi/\alpha\eta_{\ell} \approx 0.06$. Thus, altogether we find the numerical values for the cross sections,
\bea 
  \sigma^{2\to2} = 0.14\, r_{e}^{2}; \qquad \sigma^{2\to1} = 0.43\, r_{e}^2.
\eea
Therefore, for these parameters, the stimulated $2\to 1$ process in a classical background is three times more probable than the vacuum $2\to2$ process, even though the coherence factor is less than unity. We conclude that while the coherence factor is useful in quantifying the role of the classical background, it does not measure the relative importance of the stimulated compared to the vacuum process. For that, a full calculation as above is required.

\section{Conclusion}
Fundamental QED processes can be modified in coherent electromagnetic fields. When one or more of the incoming or outgoing photons are replaced by interactions with the field, the resulting `field-assisted' process can be enhanced or suppressed compared to the vacuum process. The coherence of the electromagnetic background can lead to an increase in the QED cross-section akin to the phenomenon of bosonic enhancement. The outgoing phase space can be reduced due to the spacetime dependence of the electromagnetic field and the resulting modified kinematics can lead to a suppression. In this article we have considered the introduction of a form factor to include the changes in the cross-section to the vacuum QED process brought about by the electromagnetic background. For a plane-wave electromagnetic field, such as well-approximates high-energy scattering from a laser pulse, the form factor is proportional to a coherence factor, $\xi^2\Phi/\alpha\eta$, where $\xi$ is the laser intensity parameter, $\Phi$ is the phase duration of the pulse and $\eta$ is the total light-front energy parameter of the incoming particle. This coherence factor plays a role similar to the nuclear charge in Delbr\"{u}ck scattering of photons in the Coulomb field of nuclei. Parameters of the laser beam remain in the cross-section and hence universality of the cross-section is broken when the laser beam is associated with the out state, i.e., when the laser beam is scattered \emph{into}.

If the intensity is high enough, the coherence factor acquires corrections through higher powers of $\xi^{2n}$, which at even higher intensities must be resummed and can lead to a suppression. If the field is a weakly-focussed pulse instead of a plane wave, the form factor acquires corrections of the order $\sim |\Delta \vkap|/|p_{\trm{in}}|$ where $\Delta \vkap$ is the momentum spread of the background and $p_{\trm{in}}$ is the incident particle momentum.
\newline 

The cross-section for the field-assisted process in general has a different scaling with the centre-of-mass (CM) energy; it is \emph{not} obtained by simply replacing the photon momentum in the cross-section of the vacuum process with the momentum of photons in the field. For instance, at the planned BIREF@HIBEF experiment \cite{Ahmadiniaz:2024xob}, the CM energy is $\omega=116\,\trm{eV}$. The total cross-section for the vacuum $2\to2$ and the helicity-flipping part of the field-assisted $1\to1$ process are:
\bea
  \sigma^{2\to 2} &=& \frac{973\,\alpha^{2}r_{e}^{2}}{10125 \pi}\left(\frac{\omega}{m}\right)^{6} \approx 1.8\times 10^{-53}\,\trm{m}^2  \\[5pt]
  \sigma^{1\to1}_{+-} &=& \sqrt{\frac{2}{\pi}}\frac{\xi^{2}\Phi}{\alpha} \left(\frac{2\alpha r_{e}}{15}\right)^{2}\left(\frac{\omega}{m}\right)^{4} \approx 0.9 \times 10^{-43}\,\trm{m}^{2} \nn \\
\eea
where beam parameters have been taken from \cite{Ahmadiniaz:2024xob,Macleod:2024jxl}. In our notation they correspond to $\xi \approx 30$, $\Phi \approx 40$, giving a huge coherence factor of around $\xi^{2}\Phi/\alpha\eta \approx 5 \times 10^{13}$ (recall $\eta=2\omega^{2}/m^{2}$). This coherent enhancement, together with the more favourable energy scaling of $\eta^4$ explains why the coherent cross section, $\sigma^{1\to1}_{+-}$, exceeds the vacuum light-by-light cross section, $\sigma^{2\to 2}$, by about 10 orders of magnitude! It also implies that experiments testing vacuum birefringence, which probe the $1\to1$ cross-section are currently more feasible than those testing real photon-photon scattering \cite{Inada:2014srv,Yamaji:2016xws,Watt:2024brh}, which probe the $2\to2$ cross-section.
\newline

Finally, for pair-annihilation to two photons, we have shown that the coherence factor can again lead to an enhancement of the field-assisted process compared to the vacuum process. However, due to the restrictive $2\to 1$ kinematics of the field-assisted process, realising this enhancement would require good control of the collision angle of the electron and positron. We suggest that such control is possible through the use of x-ray free electron lasers (XFELs) to assist QED processes with kinematic restrictions such as absorption and annihilation.

\section*{Acknowledgments}
The authors acknowledge support from The Leverhulme Trust, Grants RPG-2023-285 and RPG-2024-142.

\appendix

\bibliographystyle{apsrev}
\bibliography{current}

\end{document}